\documentclass{aa}

\usepackage{graphicx}
\usepackage{txfonts}
\usepackage{xcolor}
\usepackage{upgreek}
\usepackage{natbib}
\usepackage{adjustbox}
\usepackage[colorlinks,citecolor=blue,urlcolor=blue,filecolor=blue,linkcolor=blue]{hyperref}

\begin{document}
    \title{The LOFAR Tied-Array All-Sky Survey: Timing of 35 radio pulsars and an overview of the properties of the LOFAR pulsar discoveries}

    \author{
    E.\ van der Wateren\inst{\ref{astron}, \ref{nijmegen}}
    \fnmsep\thanks{\url{emma.vanderwateren@gmail.com}}
    \and
    C.\ G.\ Bassa\inst{\ref{astron}}
    \and
    S.\ Cooper \inst{\ref{jb}}
    \and
    J.\ -M.\ Grie{\ss}meier \inst{\ref{orleans}, \ref{nancay}}
    \and
    B.\ W.\ Stappers\inst{\ref{jb}}
    \and
    J.\ W.\ T.\ Hessels\inst{\ref{astron}, \ref{api}}
    \and
    V.\ I.\ Kondratiev\inst{\ref{astron}}
    \and 
    D.\ Michilli \inst{\ref{mit},\ref{kavli}}
    \and
    C.\ M.\ Tan\inst{\ref{mcgill1},\ref{mcgill2}}
    \and
    C.\ Tiburzi \inst{\ref{cagliari}}
    \and
    P.\ Weltevrede\inst{\ref{jb}}
    \and
    A.\ -S.\ Bak Nielsen \inst{\ref{mpi},\ref{bielefeld}}
    \and
    T.\ D.\ Carozzi \inst{\ref{onsala}}
    \and
    B.\ Ciardi \inst{\ref{garching}}
    \and
    I.\ Cognard\inst{\ref{orleans}, \ref{nancay}}
    \and 
    R.\ -J.\ Dettmar\inst{\ref{ruhr}}
    \and
    A.\ Karastergiou \inst{\ref{oxford}}
    \and 
    M.\ Kramer \inst{\ref{jb}, \ref{mpi}}
    \and
    J.\ K{\"u}nsem{\"o}ller\inst{\ref{bielefeld}}
    \and
    S.\ Os{\l}owski \inst{\ref{manly}}
    \and
    M.\ Serylak \inst{\ref{ska},\ref{cape}}
    \and
    C.\ Vocks \inst{\ref{leibniz}}
    \and
    O.\ Wucknitz \inst{\ref{mpi}}
    } 
   
    \institute{
        ASTRON, Netherlands Institute for Radio Astronomy, Oude Hoogeveensedijk 4, 7991 PD Dwingeloo, The Netherlands\label{astron}
        \and 
        Department of Astrophysics/IMAPP, Radboud University Nijmegen, P.O. Box 9010, 6500 GL Nijmegen, The Netherlands\label{nijmegen}
        \and
        Jodrell Bank Centre for Astrophysics, Department of Physics and Astronomy, University of Manchester, Manchester M13 9PL, UK\label{jb}
        \and
        LPC2E - Universit\'{e} d'Orl\'{e}ans/CNRS, France\label{orleans}
        \and
        ORN, Observatoire de Paris, Universit\'{e} PSL, Univ Orl\'{e}ans, CNRS, 18330 Nan\c cay, France\label{nancay}  
        \and 
        Anton Pannekoek Institute for Astronomy, University of Amsterdam, Science Park 904, 1098\,XH Amsterdam, The Netherlands\label{api}
        \and
        Department of Physics, Massachusetts Institute of Technology, 77 Massachusetts Ave, Cambridge, MA 02139, USA \label{mit}
        \and
        MIT Kavli Institute for Astrophysics and Space Research, Massachusetts Institute of Technology, 77 Massachusetts Ave, Cambridge, MA 02139, USA \label{kavli}
        \and
        Department of Physics, McGill University, 3600 rue University, Montr\'{e}al, QC H3A 2T8, Canada\label{mcgill1}
        \and
        McGill Space Institute, McGill University, 3550 rue University, Montr\'{e}al, QC H3A 2A7, Canada\label{mcgill2}
        \and
        INAF - Cagliari Astronomical Observatory, Via della Scienza 5, I–09047 Selargius (CA), Italy \label{cagliari}
        \and
        Max-Planck-Institut f{\"u}r Radioastronomie, Auf dem H{\"u}gel 69, D-53121 Bonn, Germany Bonn, Germany \label{mpi}
        \and
        Fakult{\"a}t f{\"u}r Physik, Universit{\"a}t Bielefeld, Postfach 100131, 33501 Bielefeld, Germany\label{bielefeld}
        \and
        Department of Space, Earth and Environment, Chalmers University of Technology, Onsala Space Observatory, 439 92 Onsala, Sweden \label{onsala} 
        \and
        Max-Planck-Institut f{\"u}r Astrophysik, Karl-Schwarzschild-Str. 1, 85741 Garching, Germany \label{garching}
        \and
        Ruhr University Bochum, Faculty of Physics and Astronomy, Astronomical Institute, 44780 Bochum, Germany \label{ruhr}
        \and 
        Oxford Astrophysics, Denys Wilkinson Building, Keble Road, Oxford, OX1 3RH, UK\label{oxford}
        \and
        Manly Astrophysics, 15/41-42 East Esplanade, Manly, NSW 2095, Australia \label{manly}
        \and
        SKA Observatory, Jodrell Bank, Lower Withington, Macclesfield, SK11 9FT, United Kingdom \label{ska}
        \and 
        Department of Physics and Astronomy, University of the Western Cape, Bellville, Cape Town, 7535, South Africa \label{cape}
        \and
        Leibniz-Institut f{\"u}r Astrophysik Potsdam (AIP), An der Sternwarte 16, 14482 Potsdam, Germany \label{leibniz}
    }
    \titlerunning{Timing of 35 LOTAAS pulsars}
    \authorrunning{E.\ van der Wateren et al.}


    \abstract{The LOFAR Tied-Array All-Sky Survey (LOTAAS) is the most sensitive untargeted radio pulsar survey performed at low radio frequencies (119--151\,MHz) to date and has discovered 76 new radio pulsars, among which the 23.5-s pulsar J0250+5854, up until recently the slowest-spinning radio pulsar known. Here, we report on the timing solutions of 35 pulsars discovered by LOTAAS, which include a nulling pulsar and a mildly recycled pulsar, and thereby complete the full timing analysis of the LOTAAS pulsar discoveries. We give an overview of the findings from the full LOTAAS sample of 76 pulsars, discussing their pulse profiles, radio spectra and timing parameters. We found that the pulse profiles of some of the pulsars show profile variations in time or frequency and while some pulsars show signs of scattering, a large majority display no pulse broadening. The LOTAAS discoveries have on average steeper radio spectra and have longer spin periods ($1.4\times$) as well as lower spin-down rates ($3.1\times$) compared to the known pulsar population. We discuss the cause of these differences, and attribute them to a combination of selection effects of the LOTAAS survey as well as previous pulsar surveys, though can not rule out that older pulsars tend to have steeper radio spectra.
    }


\maketitle

\section{Introduction}
Radio pulsars are uniquely useful sources to study the Galactic neutron star population, because their radio emission holds information about their rotation and relative motion, as well as the signal propagation. These properties, which are mostly unknown for other types of neutron stars, can be used to derive parameters which are essential to describe the population, such as the characteristic age, the magnetic field, and the spin-down energy \citep{Lorimer12}. Models for population synthesis rely on information about the currently known population. So far, the ATNF Pulsar Catalogue\footnote{\url{http://www.atnf.csiro.au/research/pulsar/psrcat}} (version 1.67, \citealt{Manchester05}), consists of 3320 pulsars \citep{Manchester05}, which is expected to be only a small proportion of the full Galactic pulsar population \citep{Keane15}. Surveys searching for new pulsars and exploring new parameter spaces are therefore necessary to get a complete picture of the full Galactic neutron star population. Recent discoveries resulting from radio pulsar surveys include the double pulsar \citep{Burgay03, Lyne04, Kramer21}, the fastest and heaviest Galactic neutron star \citep{Bassa17, Romani22}, and unusually long-period radio sources \citep{Hurley-Walker22, Caleb22}.

A large proportion of pulsars have been discovered in surveys at observing frequencies around 1.4\,GHz and by predominantly surveying the Galactic plane, such as the Parkes multi-beam pulsar survey \citep{Manchester01} and the PALFA survey \citep{Cordes06, Lazarus15}. More recently, there have been several wide area pulsar surveys at lower observing frequencies, including the Green Bank North Celestial Cap (GBNCC) survey, observing at 350\,MHz \citep{Stovall14}, AO327, a survey with Arecibo at 327\,MHz \citep{Deneva13}, the Giant Metrewave Radio Telescope (GMRT) High Resolution Southern Sky Survey at 322\,MHz \citep{Bhattacharyya16}, and the Pushchino Multibeam Pulsar Search (PUMPS) at 111\,MHz \citep{Tyul'bashev22}. As the currently known pulsar population is a reflection of the surveys with which they were found, searching at different frequencies may shed light on new pulsar populations.

Pulsars, which were originally discovered at the low observing frequency of 81.5\,MHz \citep{Hewish68}, have generally steep radio spectra ($S_\nu\propto\nu^{-1.57\pm0.62}$; \citealt{Jankowski18}) and hence are brighter at lower frequencies. Therefore, many pulsars may still be waiting to be discovered, being only observable at lower frequencies. However, pulsar surveys at low observing frequencies are challenging because these observations are more severely affected by dispersion and interstellar scattering, which scale with frequency $\nu$ as $\nu^{-2}$ and  $\nu^{-4}$, respectively. Additionally, because of the $\nu^{-2.6}$ sky temperature spectrum \citep[e.g.][]{Price21}, low frequency observations also suffer more from high sky temperatures. For these reasons, pulsar surveys at low frequencies are less sensitive to pulsars in the Galactic plane, where dispersion, scattering and sky temperatures are all higher compared to higher Galactic latitudes.

The LOFAR Tied-Array All-Sky Survey (LOTAAS) was aimed at searching for radio pulsars and fast transients at very low observing frequencies. This untargeted all-sky survey of the northern hemisphere operated at frequencies between 119 and 151\,MHz with long 1-hour integration times per pointing \citep{Sanidas19}. Full timing solutions and spectra of 41 pulsars discovered by LOTAAS have been presented by \citet{Tan18}, \citet{Michilli20}, and \citet{Tan20}, which include the 23.5-s pulsar J0250+5854, up until recently the slowest spinning radio pulsar known. Other discoveries were a mildly recycled binary pulsar and a Rotating Radio Transient (RRAT; \citealt{McLaughlin06}). 

By investigating the pulsars discovered by LOTAAS at very low observing frequencies, we gain a unique insight into a new parameter space of the pulsar population. In this paper, we report on the properties of 32 pulsars  discovered by LOTAAS \citep{Sanidas19} and two pulsars discovered by its pilot survey, the LOFAR Tied-Array Survey (LOTAS; \citealt{Coenen14}). We also included PSR\,J1958+2214, which was discovered as PSR\,J1958+21 as part of the LOTAAS survey after the publication of the overview paper by \citet{Sanidas19}. In \S\,\ref{sec:obs_and_ana}, we combine the LOFAR observations with observations from other telescopes to perform a multi-frequency analysis, hence obtaining the full timing solutions of all 35 pulsars. The pulse profiles at different frequencies were inspected for signs of interstellar scattering and profile variation. Additionally, we calculated flux densities and performed spectral analysis to constrain the spectral index. As this paper completes the full timing analysis of all pulsars discovered by LOTAAS, we also give an overview of the findings of all LOTAAS pulsar discoveries in \S\,\ref{sec:res}, and discuss their properties compared to the known pulsar population in \S\,\ref{sec:disc}. 

\section{Observations and analysis}
\label{sec:obs_and_ana}
\subsection{Observations}
\label{sec:obs}
Timing observations of the LOTAAS pulsars were obtained with the LOFAR Core, the International LOFAR stations in Germany, France, Sweden, and the United Kingdom, the Lovell telescope at Jodrell Bank Observatory (JBO) in the United Kingdom, and the Nan\c cay Radio Telescope in France. For each pulsar, we used the best-known celestial position, spin period, and dispersion measure ($\mathrm{DM}$) from their discovery or follow-up gridding observations \citep{Coenen14, Sanidas19} to point the telescopes and obtain timing observations.

For the LOFAR Core observations, the high-band antennas (HBAs) from the 24 Core stations were digitally beamformed with the \textsc{cobalt} correlator/beamformer \citep{Broekema18} to create a single tied-array beam pointing towards the pulsar (see \citealt{Stappers11} for a description of LOFAR's beam-formed modes). Nyquist sampled, dual polarisation complex voltages were recorded for 400 subbands of 0.1953125\,MHz each, covering observing frequencies from 110 to 188\,MHz. The complex voltages were coherently dedispersed and folded using \textsc{dspsr} \citep{VanStraten11} through the LOFAR pulsar pipeline \citep{Kondratiev16} to produce folded pulse profiles in the \textsc{psrfits} format for 10-s sub-integrations and 0.1953125\,MHz channels. We used \textsc{clfd} \citep{Morello19} to automatically remove the majority of the radio frequency interference (RFI). Remaining residual RFI was removed manually with the \textsc{psrzap} tool from the \textsc{psrchive} software suite \citep{Hotan04}.

The observations from the LOFAR stations of the German LOng Wavelength consortium (GLOW) were processed following the procedure described in \citet{Donner20}. For the observations of the French LOFAR station FR606, voltage data was recorded between 102.34 and 197.46\,MHz in subbands of 0.1953125\,MHz and processed with a pipeline based on \textsc{dspsr} producing folded pulse profiles in the \textsc{psrfits} format for 10-s sub-integrations. Frequencies below 109.96\,MHz and above 187.89\,MHz were discarded. RFI was automatically removed using \textsc{CoastGuard} \citep{Lazarus16}, followed by a manual inspection and RFI removal where needed. The pre-processing of the observations from the Swedish LOFAR station SE607 was done similarly to the FR606 procedure, with the additional steps of removing data between 181 and 186\,MHz and injecting improved ephemeris before cleaning the observations. The UK LOFAR station at Chilbolton (UK608) uses the ARTEMIS\footnote{Advanced Radio Transient Event Monitor and Identification System} backend \cite{Karastergiou15}. Observations were typically 1\,h in duration, spanned 48\,MHz, centred at 162\,MHz, and divided in to channels of approximately 12.2\,kHz. The data were folded with a pipeline based on \textsc{dspsr} and further analysis was done using \textsc{psrchive}. The majority of RFI was automatically removed using \textsc{clfd} and the remainder was removed manually with \textsc{psrzap}

Observations with the LOFAR Core were obtained at an approximately monthly cadence, while the International LOFAR stations observed at an approximately weekly or semi-weekly cadence. Integration times of the LOFAR Core observations were between 10 and 20\,min while the lower-collecting-area International LOFAR stations observed between 1 and 4\,hr. 

At higher observing frequencies, the majority of the LOTAAS pulsars were observed with the Lovell telescope at 334\,MHz and 1532\,MHz to constrain their spectra. For those pulsars detected at 1532\,MHz, further timing observations were obtained at a weekly to monthly cadence. These observations were obtained using the Digital Filterbank (DFB) with observed bandwidths of 64\,MHz at 334\,MHz, and 384\,MHz at 1532\,MHz. The integration times were between a few minutes and an hour depending on the pulsar brightness. The Nan\c cay Radio Telescope observed at 1484\,MHz with the NUPPI backend \citep{Guillemot16}. The integration times were between a few minutes and an hour with a bandwidth of 512\,MHz. As only a few JBO observations at 334\,MHz and NRT observations were obtained per pulsar, these were not used for timing, but only to determine flux densities and pulse widths. An overview of all observations is provided in Table\,\ref{tab:obs1}.

\begin{table}[]
    \caption{Overview of the radio observations. Pulsar names are provided based on the timing positions, as well as the discovery designation with which they were reported in \citet{Coenen14} and \citet{Sanidas19} (disc.). The observation time span is given by $t_\mathrm{span}$ and the number of observations from LOFAR Core (c), International Stations (i) the Lovell telescope at 334\,MHz (P-band), 1532\,MHz (L-band) and the Nan\c cay telescope at 1484\,MHz (L-band) is as indicated.}
    \centering
    \tiny
    \begin{tabular}{llrrrrrr}
    \hline
        Pulsar & Pulsar  & \multicolumn{1}{l}{$t_\mathrm{span}$} & \multicolumn{5}{c}{$N_\mathrm{obs}$} \\
        & (disc.) &(yr) & \multicolumn{2}{c}{LOFAR} & \multicolumn{2}{c}{Lovell} & NRT \\
        &  &    & (c) & (i) & (P) & (L) & (L) \\
    \hline
J0039+3546   & J0039+35   & 6.5 & 34 & 242 & 1 & 8 & 9\\ 
J0059+6956   & J0059+69   & 2.4 & 33 & & & & 3 \\
J0139+5621   & J0140+5622 & 7.9 & 50 & & 2 & 57 & \\
J0305+1123   & J0305+11   & 5.1 & 67 & & 1 &7 & \\
J0317+1328   & J0317+13   & 3.1 & 37 & & 1 & 3 & 1 \\
J0613+3731   & J0613+3731 & 9.0 & 56 & 416 & 3 & 84 & \\
J0811+3729   & J0811+37   & 4.1 & 43 & & 2 & 6 & 3\\
J0813+2202   & J0813+22   & 2.0 & 29 & & & 11 & 5\\
J0828+5304   & J0827+53   & 2.4 & 64 & & & 24 & 3\\
J0928+3039   & J0928+30   & 2.6 & 31 & & 1 & 6 & 5\\
J0935+3312   & J0935+33   & 3.9 & 39 & 50 & 3 & 6 & 10\\
J1303+3815   & J1303+38   & 1.7 & 24 & & & 17 & 3\\
J1334+1005   & J1334+10   & 1.7 & 24 & & & & 5 \\
J1427+5211   & J1426+52   & 2.4 & 34 & & & 14 & 5\\
J1529+4050   & J1529+40   & 7.1 & 38 & 203 & 2 & 104 & 4\\
J1707+3556   & J1707+35   & 1.9 & 25 & & & & 3\\
J1715+4603   & J1715+46   & 2.5 & 32 & & 1 & 29 & 3\\
J1722+3519   & J1722+35   & 2.4 & 28 & & 1& 5 & 4\\
J1740+2715   & J1740+27   & 6.0 & 32 & 210 & 1 &52 & 3\\
J1745+4254   & J1745+42   & 5.8 & 52 & & 1 & 49 & 2\\
J1809+1705   & J1809+17   & 3.0 & 41 & & 1 & 8 & 3\\
J1814+2224   & J1814+22   & 6.8 & 33 & 258 & 1 & 77 & 1\\
J1910+5655   & J1910+56   & 3.0 & 53 & & & 18 & 4\\
J1953+3014   & J1953+30   & 2.4 & 34 & & & 46 & 4 \\
J1958+2214   & J1958+21   & 2.2 & 26 & & & & \\
J1958+5650   & J1958+56   & 2.3 & 34 & & & 37 & 1 \\
J2006+2205   & J2006+22   & 2.7 & 32 & & 1& 5& 6\\
J2022+2112   & J2022+21   & 2.0 & 28 & & & 5 & 3\\
J2053+1718   & J2053+17   & 2.5 & 32 & & & 5 & 4\\
J2057+2128   & J2057+21   & 3.1 & 42 & & & & 6\\
J2123+3624   & J2123+36   & 2.4 & 33 & & & 15 & 3\\
J2209+2117   & J2209+22   & 6.2 & 33 & & & 121 & 3\\
J2306+3124   & J2306+31   & 6.2 & 28 & 213 & &140 & 3\\
J2336$-$0151 & J2336$-$01 & 3.1 & 43 & & 1 & 4 & 9\\
J2350+3140   & J2350+31   & 3.6 & 36 & 38 &1 & 7 & 14\\
		\hline
    \end{tabular}
    \label{tab:obs1}
\end{table}

\begin{table*}
    \tiny
    \caption{Pulsar timing solutions.}
    \begin{tabular}{llllllllll}
        \hline
        PSR & $\alpha_\mathrm{J2000}$ & $\delta_\mathrm{J2000}$ & Epoch & $P$ & $\dot{P}$ & $\mathrm{DM}$ & $\mathrm{N_{TOA}}$ & $\mathrm{RMS}$ & $\mathrm{\chi^2_{red}}$\\
         & & & (MJD) & ($\mathrm{s}$) & ($\mathrm{10^{-15}}$) & ($\mathrm{pc \, cm^{-3}}$) &  & ($\upmu$s) & \\
        \hline
        J0039+3546 & $00^\mathrm{h}39^\mathrm{m}08\fs8171(6)$ & $+35\degr46\arcmin16\farcs432(10)$ & 58461 & 0.53668651031180(15) & 0.063385(4) & 53.0249(3) & 244 & 139 & 1.36 \\
        J0059+6956 & $00^\mathrm{h}59^\mathrm{m}39\fs617(3)$ & $+69\degr56\arcmin31\farcs34(2)$ & 58556 & 1.1459097060823(13) & 0.14070(15) & 63.4768(5) & 33 & 127 & 0.74 \\
        J0139+5621 & $01^\mathrm{h}39^\mathrm{m}38\fs577(16)$ & $+56\degr21\arcmin36\farcs38(16)$ & 57835 & 1.775356366845(11) & 79.1067(4) & 101.83(4) & 102 & 1954 & 8.01 \\
        J0305+1123 & $03^\mathrm{h}05^\mathrm{m}07\fs291(6)$ & $+11\degr23\arcmin24\farcs9(3)$ & 58055 & 0.862062679407(4) & 0.30478(12) & 27.830(17) & 67 & 455 & 1.41 \\
        J0317+1328 & $03^\mathrm{h}17^\mathrm{m}48\fs66(2)$ & $+13\degr28\arcmin34\farcs0(1.2)$ & 57673 & 1.974239666372(14) & 0.1696(9) & 12.57(18) & 37 & 885 & 2.31 \\
         J0613+3731 & $06^\mathrm{h}13^\mathrm{m}12\fs137(3)$ & $+37\degr31\arcmin38\farcs53(13)$ & 58128 & 0.6191987768316(6) & 3.237300(16) & 18.9677(10) & 544 & 1279 & 797.45 \\
        J0811+3729 & $08^\mathrm{h}11^\mathrm{m}15\fs095(13)$ & $+37\degr29\arcmin13\farcs8(6)$ & 58238 & 1.248264327889(10) & 0.7458(5) & 17.100(4) & 34 & 1003 & 19.02 \\
        J0813+2202 & $08^\mathrm{h}13^\mathrm{m}51\fs94(2)$ & $+22\degr02\arcmin16\farcs3(1.3)$ & 58626 & 0.5313827887886(19) & 0.0777(2) & 52.2555(20) & 29 & 333 & 0.83 \\
        J0828+5304 & $08^\mathrm{h}28^\mathrm{m}25\fs69737(20)$ & $+53\degr04\arcmin43\farcs053(3)$ & 59065 & 0.0135273958162361(13) & $2.597(9)\times10^{-5}$ & 23.10340(6) & 60 & 16 & 1.7 \\
        J0928+3039 & $09^\mathrm{h}28^\mathrm{m}59\fs360(9)$ & $+30\degr39\arcmin26\farcs5(3)$ & 57270 & 1.045756634409(18) & 0.3629(4) & 22.3(3) & 31 & 628 & 14.09 \\
        J0935+3312 & $09^\mathrm{h}35^\mathrm{m}07\fs799(4)$ & $+33\degr12\arcmin36\farcs62(13)$ & 57512 & 0.961553419242(3) & 0.37797(15) & 18.33(2) & 87 & 548 & 1.49 \\
        J1303+3815 & $13^\mathrm{h}03^\mathrm{m}19\fs3477(5)$ & $+38\degr15\arcmin03\farcs167(7)$ & 58681 & 0.39627397772729(20) & 0.34938(3) & 18.99746(14) & 23 & 31 & 1.6 \\
        J1334+1005 & $13^\mathrm{h}34^\mathrm{m}32\fs626(13)$ & $+10\degr05\arcmin41\farcs5(4)$ & 58688 & 0.911092162286(13) & 0.7415(16) & 23.868(3) & 24 & 780 & 0.85 \\
        J1427+5211 & $14^\mathrm{h}27^\mathrm{m}07\fs711(3)$ & $+52\degr11\arcmin11\farcs96(2)$ & 58555 & 0.995864060026(2) & 0.4744(2) & 25.3107(9) & 34 & 205 & 1.48 \\
        J1529+4050 & $15^\mathrm{h}29^\mathrm{m}16\fs526(6)$ & $+40\degr50\arcmin57\farcs79(6)$ & 58032 & 0.4764184320778(13) & $2.65(3)\times10^{-3}$ & 6.484(3) & 330 & 1767 & 3.21 \\
        J1707+3556 & $17^\mathrm{h}07^\mathrm{m}02\fs7328(3)$ & $+35\degr56\arcmin36\farcs522(4)$ & 58656 & 0.15976470098275(6) & $8.52(7)\times10^{-4}$ & 19.23798(11) & 25 & 24 & 3.36 \\
        J1715+4603 & $17^\mathrm{h}15^\mathrm{m}43\fs770(7)$ & $+46\degr03\arcmin59\farcs15(7)$ & 57782 & 0.548097027876(2) & 0.0484(3) & 19.767(2) & 31 & 527 & 0.68 \\
        J1722+3519 & $17^\mathrm{h}22^\mathrm{m}09\fs5067(6)$ & $+35\degr19\arcmin18\farcs591(8)$ & 57795 & 0.8216175744405(4) & 0.54474(5) & 23.8766(2) & 27 & 54 & 4.72 \\
        J1740+2715 & $17^\mathrm{h}40^\mathrm{m}32\fs5712(10)$ & $+27\degr15\arcmin21\farcs470(16)$ & 58300 & 1.0582099056193(11) & 0.21430(3) & 35.522(5) & 281 & 348 & 1.54 \\
        J1745+4254 & $17^\mathrm{h}45^\mathrm{m}50\fs1197(13)$ & $+42\degr54\arcmin37\farcs895(17)$ & 58333 & 0.3050545538470(2) & $9.851(8)\times10^{-3}$ & 37.931(9) & 92 & 227 & 1.0 \\
        J1809+1705 & $18^\mathrm{h}09^\mathrm{m}07\fs483(7)$ & $+17\degr05\arcmin44\farcs54(17)$ & 57663 & 2.066649687618(13) & 3.0727(12) & 47.187(3) & 38 & 951 & 1.14 \\
        J1814+2224 & $18^\mathrm{h}14^\mathrm{m}29\fs9046(5)$ & $+22\degr24\arcmin40\farcs183(10)$ & 58153 & 0.25371005873953(6) & 0.354845(2) & 62.2999(19) & 357 & 230 & 1.14 \\
        J1910+5655 & $19^\mathrm{h}10^\mathrm{m}53\fs64(3)$ & $+56\degr55\arcmin33\farcs5(2)$ & 58980 & 0.341858932039(4) & 0.2048(2) & 20.658(5) & 51 & 2306 & 7.45 \\
        J1953+3014 & $19^\mathrm{h}53^\mathrm{m}52\fs445(2)$ & $+30\degr14\arcmin30\farcs63(3)$ & 58555 & 1.271207766231(3) & 0.3852(3) & 43.5777(11) & 34 & 225 & 1.48 \\
        J1958+2214 & $19^\mathrm{h}58^\mathrm{m}45\fs284(9)$ & $+22\degr14\arcmin26\farcs11(18)$ & 59299 & 1.050380009903(12) & 4.5748(13) & 87.665(4) & 26 & 927 & 0.65 \\
        J1958+5650 & $19^\mathrm{h}58^\mathrm{m}06\fs981(6)$ & $+56\degr50\arcmin07\farcs57(4)$ & 58570 & 0.3118323776425(10) & 0.17214(12) & 58.1231(14) & 33 & 364 & 0.7 \\
        J2006+2205 & $20^\mathrm{h}06^\mathrm{m}44\fs802(5)$ & $+22\degr05\arcmin26\farcs06(10)$ & 57736 & 1.741873881197(8) & 5.8099(7) & 130.4831(19) & 30 & 549 & 1.78 \\
        J2022+2112 & $20^\mathrm{h}22^\mathrm{m}27\fs649(3)$ & $+21\degr12\arcmin38\farcs64(7)$ & 58626 & 0.803551296937(3) & 0.9048(4) & 73.5194(14) & 28 & 334 & 1.86 \\
        J2053+1718 & $20^\mathrm{h}53^\mathrm{m}49\fs4814(2)$ & $+17\degr18\arcmin44\farcs731(6)$ & 57765 & 0.11926776027515(3) & 0.000301(3) & 26.97952(18) & 31 & 24 & 3.87 \\
        J2057+2128 & $20^\mathrm{h}57^\mathrm{m}38\fs237(14)$ & $+21\degr28\arcmin05\farcs4(3)$ & 57669 & 1.166639603609(12) & 0.5229(10) & 73.12(17) & 40 & 1679 & 0.9 \\
        J2123+3624 & $21^\mathrm{h}23^\mathrm{m}58\fs407(12)$ & $+36\degr24\arcmin39\farcs30(19)$ & 58556 & 1.294029820771(13) & 2.1008(14) & 108.525(4) & 30 & 1069 & 1.25 \\
        J2209+2117 & $22^\mathrm{h}09^\mathrm{m}59\fs7239(11)$ & $+21\degr17\arcmin58\farcs49(2)$ & 58335 & 1.776960679496(5) & 5.07777(11) & 46.3741(4) & 104 & 213 & 1.69 \\
        J2306+3124 & $23^\mathrm{h}06^\mathrm{m}19\fs2039(7)$ & $+31\degr24\arcmin20\farcs364(13)$ & 58234 & 0.34160250343014(13) & 0.222456(4) & 46.15171(20) & 307 & 219 & 2.8 \\
        J2336$-$0151 & $23^\mathrm{h}36^\mathrm{m}32\fs07(17)$ & $-01\degr51\arcmin56''.7(5.8)$ & 57669 & 1.029839015486(4) & 0.3448(3) & 19.622(13) & 39 & 630 & 0.99 \\
        J2350+3140 & $23^\mathrm{h}50^\mathrm{m}41\fs200(3)$ & $+31\degr40\arcmin47\farcs24(5)$ & 57778 & 0.5080916900926(7) & 0.35236(5) & 39.141(14) & 74 & 353 & 2.06 \\
        \hline
    \end{tabular}
    \label{tab:timing}
    \tablefoot{The spin period and spin period derivative are referenced to the quoted Epoch. Figures in parentheses are the nominal $1\sigma$ \textsc{tempo2} uncertainties in the least-significant digits quoted multiplied by the square root of the reduced $\chi^2$. For PSRs\,J0828+5304 and J2053+1718, significant measurements of proper motion were obtained.}
\end{table*}

\subsection{Pulsar timing}\label{sec:ana_timing}
For each pulsar, we aimed to obtain a phase-coherent timing solution which accounts for all rotations of the pulsar over the observing time span and allows for prediction of the arrival time of pulses as a function of time and observing frequency. As all pulsars described in this work, except PSR\,J0828+5304, are isolated, the timing models required fitting for the astrometric position (right ascension $\alpha_\mathrm{J2000}$ and declination $\delta_\mathrm{J2000}$), the spin period $P$, the spin period derivative $\dot{P}$, and the dispersion measure $\mathrm{DM}$, which we assumed to be constant. 

We used an iterative procedure to obtain phase-coherent timing solutions using tools from \textsc{psrchive} and \textsc{tempo2} \citep{Hobbs06}. First, to obtain phase-coherence over the observing time span we started with the LOFAR Core observations, folded and dedispersed at the discovery parameters \citep{Coenen14,Sanidas19}, as these have the highest signal-to-noise and generally the longest timing baselines. For each pulsar, we selected the highest signal-to-noise observation, averaged it in time and frequency, and modelled the resulting profile as the sum of several von Mises functions (with different positions, heights, and widths) using the \textsc{paas} tool to obtain an analytical template profile. Times-of-arrival (TOAs) were determined by cross-correlating the fully time- and frequency-averaged pulse profiles from LOFAR Core observations against this analytical template profile with \textsc{pat}, using the Fourier domain with Markov chain Monte Carlo algorithm \citep{Taylor92}. 

Due to the uncertainties in the discovery spin period and celestial position, the uncertainty in the predicted time of arrival between consecutive LOFAR Core observations was generally much larger than the spin period, making it {\it a priori} not possible to phase-connect consecutive observations. Instead, we used the algorithm described in \citet{Michilli20}, which folds the TOAs with \textsc{tempo2} over a range of spin periods around the discovery spin period, keeping the celestial position and DM fixed to the discovery values and the spin period derivative to zero. The resulting residual plots were inspected manually to identify spin periods which yielded phase coherence over part of the timing baseline. These spin periods were used to extend phase coherence over the entire timing baseline by iteratively fitting for and improving the spin period, the spin period derivative, and the celestial position. 

This approach yielded phase-coherent timing solutions for the majority of the isolated pulsars, except for four pulsars (PSRs\,J1745+4254, J1910+5655, J2053+1718, and J2306+3124) for which the algorithm had to be repeated for a range of celestial positions before finding phase coherence. In the case of the faint pulsar J1814+2224, neither approach yielded partial phase coherence. Instead, we used the Lovell observations at 1532\,MHz, which were obtained at a cadence of a few days during 2014, to find the initial phase-coherent timing solution, which was then improved and extended by using the LOFAR Core observations. 

For the binary pulsar J0828+5304, we also included parameters describing the binary orbit: the orbital period $P_\mathrm{b}$, the projected semi-major axis of the orbit $x$, the time of ascending node passage $T_\mathrm{asc}$, and the Laplace-Lagrange parameters $\eta=e\sin\omega$ and $\kappa=e\cos\omega$, with $e$ the eccentricity and $\omega$ the longitude of periastron. The Laplace-Lagrange parameters are from the ELL1 binary model \citep{Lange01}, which models orbits with small eccentricities. We used 13 LOFAR Core observations obtained over the course of one orbital period of about 5.9\,days to first obtain phase coherence over one orbit. Initial values for $P$, $P_\mathrm{b}$, $x$, and $T_\mathrm{asc}$ (assuming a circular orbit) were obtained by fitting a sinusoid to spin period measurements determined with \textsc{pdmp}. Using these initial values, a brute force search over all four parameters was used to fit TOAs over the same time span against a timing model with a circular orbit to find values for $P$, $P_\mathrm{b}$, $x$, and $T_\mathrm{asc}$ that yielded phase coherence and minimal residuals over a single orbit. Once found, the timing solution was manually extended to all observations by including the celestial position and spin period derivative when fitting the TOAs with \textsc{tempo2}. 

Once we had obtained initial phase coherent timing solutions for all pulsars, the observations from all telescopes were refolded using the new timing parameters and new template profiles were constructed for each telescope by fully averaging profiles from all observations in time and frequency and fitting von Mises functions to the averaged profiles. For each pulsar observed with different telescopes, the templates were aligned using the \textsc{pas} tool. In the case of PSR\,J1529+4050, its pulse profile displays two components of which the leading component diminishes towards higher frequencies. For this reason, the templates at 334 and 1532\,MHz were aligned against the second component of the template at 149\,MHz. To measure DM, all observations were averaged to two frequency channels from which TOAs were calculated. The LOFAR Core observations had centre frequencies of 129 and 166\,MHz, the GLOW observations were at 136 and 170\, MHz, the Chilbolton observations were at 150 and 174\,MHz, the FR606 and SE607 observations were at 129 and 167\,MHz, and the JBO observations were at 1447 and 1630\,MHz. The resulting TOAs were fitted for all timing parameters. Phase offsets were included in the timing solutions to take into account any remaining time delays between template profiles from different instruments. We note that for PSR\,J0811+3729, which displays nulls (see \S\,\ref{sec:profevol}), only sub-integrations which show emission were used to construct templates and calculate TOAs. 

\begin{figure*}
  \centering
  \includegraphics[width=\textwidth]{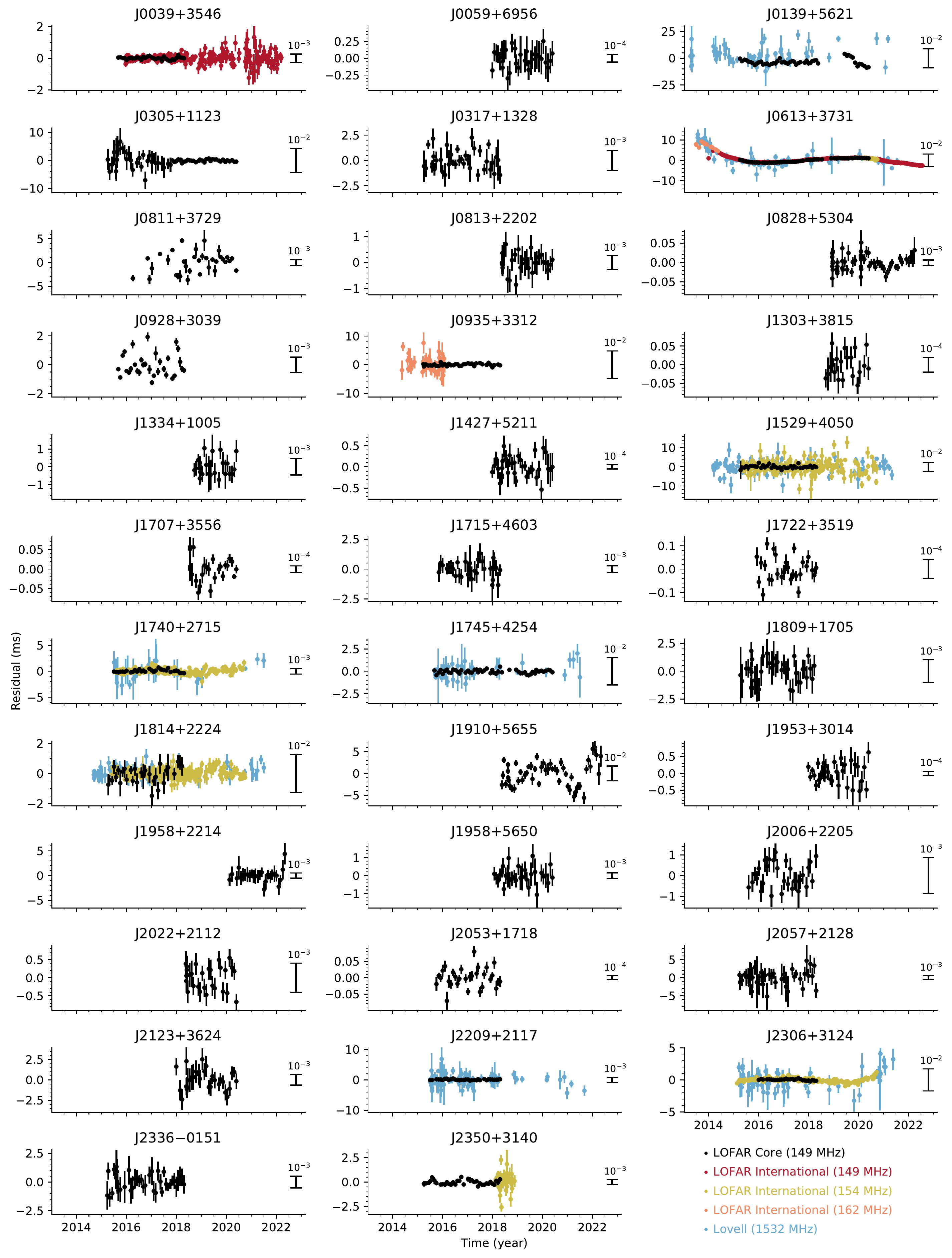}
  \caption{The residuals from the timing model in Table\,\ref{tab:timing} for all 35 pulsars. The different colours represent different instruments as indicated in the bottom-right corner. The black bars on the right side of each panel show the scale of the proportion of the spin period.}
      \label{fig:residuals}
\end{figure*}

The resulting timing solutions, now with updated DM values, were used to refold and re-dedisperse the observations, which were again fully averaged in time and frequency to determine TOAs. These TOAs were fitted for all timing parameters except DM to obtain the timing models provided in Tables\,\ref{tab:timing} and \ref{tab:J0828} and the timing residuals shown in Fig.\,\ref{fig:residuals}.  After inspection of the timing residuals of all pulsars, we also included parameters describing the proper motion of PSRs\,J0828+5304 and J2053+1718, as this improved the $\chi^2$ value of their fit and yielded significant detections of proper motion (see Section \ref{sec:res_timing}). All timing solutions were referenced to the Solar-system barycentre with the DE436 solar system ephemeris \citep{Folkner06} and the Terrestrial Time standard (BIPM2011; \citealt{Petit10}) using Barycentric Coordinate Time (TCB). 
\begin{table}
\footnotesize
\caption{Binary parameters for PSR J0828+5304.}
\begin{tabular}{ll}
\hline\hline
Orbital period, $P_b$ (d)\dotfill & 5.899255592(5) \\ 
Projected semi-major axis of orbit, $x$ (lt-s)\dotfill & 13.147610(3) \\ 
Time of ascending node passage $T_\mathrm{asc}$ (MJD)\dotfill & 58886.8001661(3) \\ 
$\eta = e \sin \omega $ \dotfill & $-$2.2(5)$\times 10^{-6}$ \\ 
$\kappa = e \cos \omega $\dotfill & $-$1.1(6)$\times 10^{-6}$ \\ 
\hline
\end{tabular}
\label{tab:J0828}
\end{table}

\subsection{Pulse profiles}
\label{sec:profiles}
The profiles at different frequencies were aligned by applying the model in Table\,\ref{tab:timing} and then rotated such that the highest peak of the LOFAR pulse profile was located at a phase $\phi=0.25$. 

\begin{figure*}
  \centering
  \includegraphics[width=\textwidth]{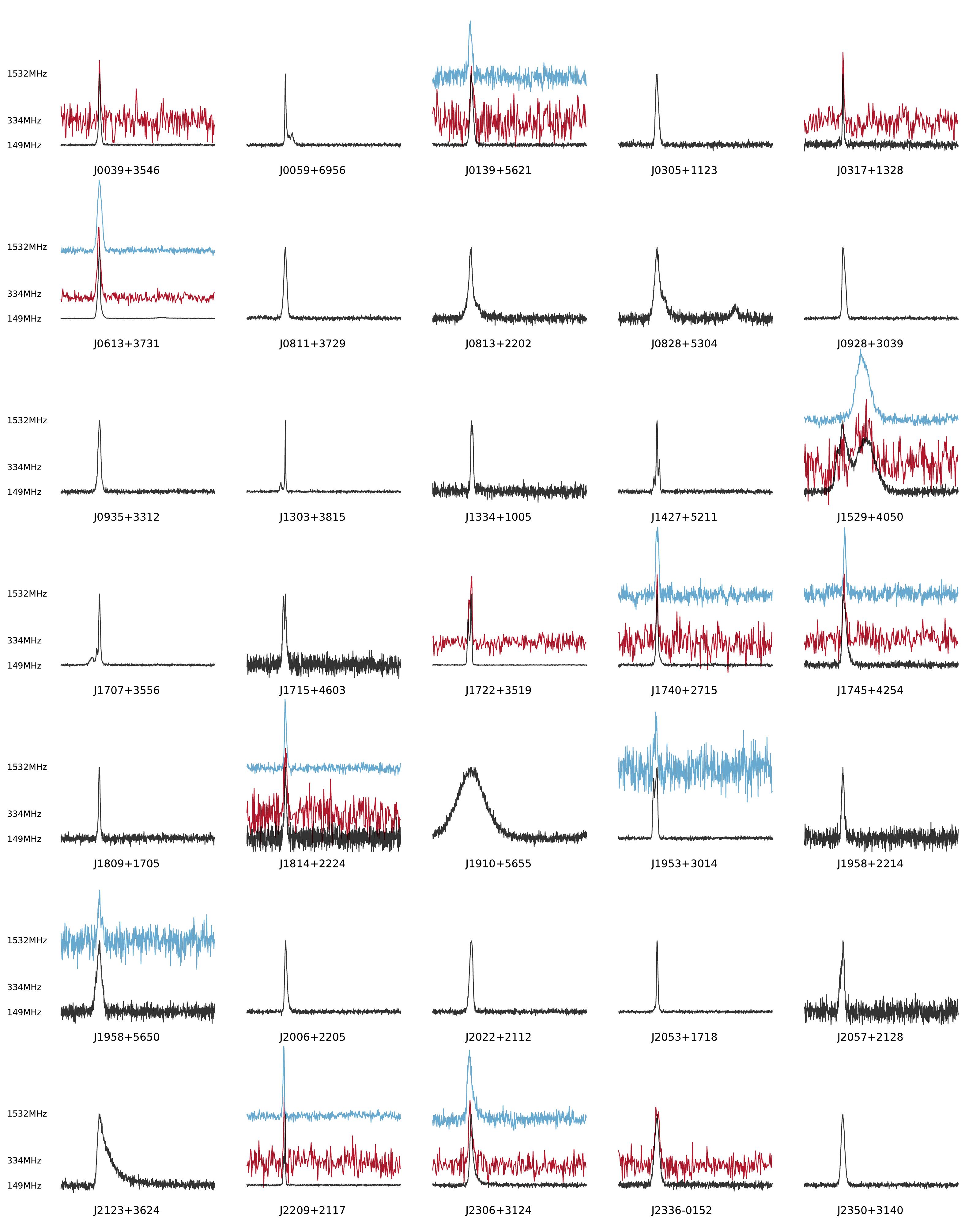}
  \caption{Pulse profiles averaged over all observations of the pulsars presented here at 149\,MHz (black), 334\,MHz (red), and 1532\,MHz (blue) with bandwidths of 78, 64 and 384\,MHz. The profiles were aligned by applying the model in Table\,\ref{tab:timing}. Then, all profiles were rotated such that the highest peak of the LOFAR pulse profile is located at phase $\phi=0.25$. The pulse profiles at 149\,MHz have 1024 phase bins, while the profiles at 334 and 1532\,MHz were rebinned from 512 and 1024 to 256 and 512 phase bins, respectively.}
    \label{fig:profiles}
\end{figure*}

The averaged pulse profiles from the pulsars presented here are shown in Fig.\,\ref{fig:profiles}. Most pulse profiles were modelled with one, two or three von Mises functions. All pulsars display a profile with a single main pulse, except for PSRs\,J0613+3731 and J0828+5304.  PSR\,J0613+3731 shows a very faint interpulse in the FR606 data, which is only just visible in the LOFAR Core data and is at a phase $\Delta\phi=0.4$ from the main pulse. The binary pulsar J0828+5304 displays a clear interpulse at $\Delta\phi=0.5$ of the main pulse.

We calculated full pulse widths at 10\% ($W_{10}$) and 50\% ($W_{50}$) of the peak intensity of the fully frequency and time averaged profiles at 149\,MHz, 334\,MHz, and 1484 or 1532\,MHz, if available. Pulse widths were determined using tools from \textsc{PSRSalsa} \citep{Weltevrede16}, which models the profiles with analytical templates composed of one or more von Mises functions and calculates pulse widths from the added components. For the observations that were used for timing, we applied the templates that were created to calculate TOAs, while for all other observations, new templates were constructed. Uncertainties on the pulse widths were determined by generating a width distribution composed of simulated pulse profiles with randomly added noise from the off-pulse region of the added profiles and then calculating the standard deviation of the simulated width distribution. The pulse widths and duty cycles can be found in Table\,\ref{tab:widths}.

\begin{sidewaystable*}
    \footnotesize
        \caption{Pulse widths and duty cycles at 10\% and 50\% of the pulse amplitude at different frequencies. Measurements with * were measured at 1484\,MHz. The uncertainties in the parentheses represent the standard deviations of simulated width distributions (see text).}
    \begin{tabular}{l | lll | lll | lll | lll}
        \hline
        PSR & \multicolumn{3}{c}{W$_{10}$ (ms)} & \multicolumn{3}{c}{$\delta_{10}$ (\%)} &\multicolumn{3}{c}{W$_{50}$ (ms)} &\multicolumn{3}{c}{$\delta_{50}$ (\%)}   \\
       & 149 & 334 & 1532  & 149 & 334 & 1532 & 149 & 334 & 1532 & 149 & 334 & 1532\\
       & MHz & MHz& MHz& MHz& MHz& MHz& MHz& MHz& MHz& MHz& MHz& MHz\\
        \hline
J0039+3546&17.6(3)&9.1(1.5)&22(4)&3.28(5) & 1.7(3) & 4.2(7) &6.95(5)&5.7(9) &17.7(1.1) & 1.296(10)&1.07(16) & 3.3(2) \\
J0059+6956&65.4(6)& $-$ & $-$ &5.71(5) &  $-$  &  $-$  &8.21(14)& $-$  & $-$  & 0.716(12)& $-$  &  $-$  \\
J0139+5621&81.3(8)&74(10)&83(3)&4.58(4) & 4.2(6) & 4.69(19) &40.2(6)&52(7) &43.5(1.8) & 2.26(3)&2.9(4) & 2.45(10) \\
J0305+1123&35.7(6)& $-$ & $-$ &4.14(7) &  $-$  &  $-$  &19.3(3)& $-$  & $-$  & 2.24(4)& $-$  &  $-$  \\
J0317+1328&34.4(6)&40(7)& $-$ &1.74(3) & 2.0(3) &  $-$  &18.9(3)&19(3) & $-$  & 0.956(17)&0.96(17) &  $-$  \\
J0613+3731&24.0(4)&29.1(8)&38.2(4)&3.87(7) & 4.70(13) & 6.17(6) &9.721(3)&15.0(4) &20.58(19) & 1.5700(5)&2.42(6) & 3.32(3) \\
J0811+3729&56.0(7)& $-$ & $-$ &4.49(6) &  $-$  &  $-$  &28.3(2)& $-$  & $-$  & 2.271(20)& $-$  &  $-$  \\
J0813+2202&52.3(1.5)& $-$ & $-$ &9.9(3) &  $-$  &  $-$  &14.1(5)& $-$  & $-$  & 2.66(9)& $-$  &  $-$  \\
J0828+5304&1.6(1.8)& $-$ & $-$ &12(13) &  $-$  &  $-$  &0.55(2)& $-$  & $-$  & 4.09(17)& $-$  &  $-$  \\
J0928+3039&43.9(4)& $-$ & $-$ &4.20(4) &  $-$  &  $-$  &25.412(16)& $-$  & $-$  & 2.4300(15)& $-$  &  $-$  \\
J0935+3312&37.8(2)& $-$ & $-$ &3.93(2) &  $-$  &  $-$  &20.56(13)& $-$  & $-$  & 2.138(13)& $-$  &  $-$  \\
J1303+3815&15.39(15)& $-$ & $-$ &3.88(4) &  $-$  &  $-$  &1.89(3)& $-$  & $-$  & 0.477(7)& $-$  &  $-$  \\
J1334+1005&28.3(1.1)& $-$ &43(3)*&3.11(12) &  $-$  & 4.7(3)* &16.8(4)& $-$  &23(2)* & 1.84(4)& $-$  & 2.6(3)* \\
J1427+5211&45.6(5)& $-$ &13.8(1.6)&4.58(5) &  $-$  & 1.38(16) &9.70(10)& $-$  &6.5(9) & 0.974(10)& $-$  & 0.65(9) \\
J1529+4050&161.6(9)&108(10)&96(5)&33.91(19) & 23(2) & 20.2(1.1) &121.0(6)&52(4) &47.7(1.0) & 25.40(14)&10.9(8) & 10.0(2) \\
J1707+3556&10.5(1.1)& $-$ & $-$ &6.6(7) &  $-$  &  $-$  &1.843(15)& $-$  & $-$  & 1.154(10)& $-$  &  $-$  \\
J1715+4603&22.3(1.0)& $-$ &31(6)&4.07(18) &  $-$  & 5.7(1.1) &15.1(5)& $-$  &20(2) & 2.76(9)& $-$  & 3.6(4) \\
J1722+3519&30.68(6)&28.0(1.4)& $-$ &3.733(7) & 3.41(17) &  $-$  &23.08(3)&18.4(1.6) & $-$  & 2.810(4)&2.24(19) &  $-$  \\
J1740+2715&37.9(7)&25(4)&42.8(1.3)&3.58(6) & 2.4(4) & 4.04(13) &14.63(12)&14.5(1.6) &23.7(7) & 1.383(11)&1.37(15) & 2.24(7) \\
J1745+4254&18.8(4)&7.6(1.0)&9.1(4)&6.15(13) & 2.5(3) & 2.98(15) &9.08(17)&3.9(3) &5.1(3) & 2.98(6)&1.29(11) & 1.67(9) \\
J1809+1705&48.3(8)& $-$ & $-$ &2.34(4) &  $-$  &  $-$  &26.8(4)& $-$  & $-$  & 1.295(19)& $-$  &  $-$  \\
J1814+2224&8.7(3)&15.3(1.6)&7.08(15)&3.44(12) & 6.0(6) & 2.79(6) &4.64(19)&4.5(8) &3.79(8) & 1.83(7)&1.8(3) & 1.50(3) \\
J1910+5655&138.0(7)& $-$ & $-$ &40.4(2) &  $-$  &  $-$  &72.1(4)& $-$  & $-$  & 21.09(10)& $-$  &  $-$  \\
J1953+3014&54.0(3)& $-$ &58(5)&4.25(3) &  $-$  & 4.6(4) &43.0(5)& $-$  &33(2) & 3.38(4)& $-$  & 2.59(17) \\
J1958+2214&41.7(1.1)& $-$ & $-$ &3.97(10) &  $-$  &  $-$  &23.2(6)& $-$  & $-$  & 2.21(6)& $-$  &  $-$  \\
J1958+5650&24.2(3)& $-$ &14.1(1.5)&7.76(11) &  $-$  & 4.5(5) &13.14(18)& $-$  &8.6(8) & 4.21(6)& $-$  & 2.7(3) \\
J2006+2205&62.3(1.1)& $-$ &49(4)*&3.58(6) &  $-$  & 2.8(3)* &28.2(4)& $-$  &26.4(1.2)* & 1.62(2)& $-$  & 1.52(7)* \\
J2022+2112&35.2(6)& $-$ & $-$ &4.39(8) &  $-$  &  $-$  &20.9(2)& $-$  & $-$  & 2.60(3)& $-$  &  $-$  \\
J2053+1718&2.53(6)& $-$ & $-$ &2.12(5) &  $-$  &  $-$  &1.138(11)& $-$  & $-$  & 0.954(9)& $-$  &  $-$  \\
J2057+2128&54(3)& $-$ & $-$ &4.6(3) &  $-$  &  $-$  &35.0(1.3)& $-$  & $-$  & 3.00(11)& $-$  &  $-$  \\
J2123+3624&250(6)& $-$ & $-$ &19.3(5) &  $-$  &  $-$  &81(5)& $-$  & $-$  & 6.3(4)& $-$  &  $-$  \\
J2209+2117&36.45(19)& $-$ &31.7(1.4)&2.051(11) &  $-$  & 1.78(8) &9.26(14)& $-$  &16.6(8) & 0.521(8)& $-$  & 0.93(4) \\
J2306+3124&22.8(4)& $-$ &30.6(2.0)&6.67(10) &  $-$  & 9.0(6) &7.07(8)& $-$  &10.0(5) & 2.07(2)& $-$  & 2.93(15) \\
J2336-0152&70.2(5)&61(5)& $-$ &6.81(5) & 6.0(5) &  $-$  &38.5(3)&36(3) & $-$  & 3.73(2)&3.5(3) &  $-$  \\
J2350+3140&25.9(3)& $-$ & $-$ &5.09(7) &  $-$  &  $-$  &13.36(16)& $-$  & $-$  & 2.63(3)& $-$  &  $-$  \\
        \hline
    \end{tabular}
    \label{tab:widths}
\end{sidewaystable*}

\begin{figure*}
  \centering
  \includegraphics[width=\textwidth]{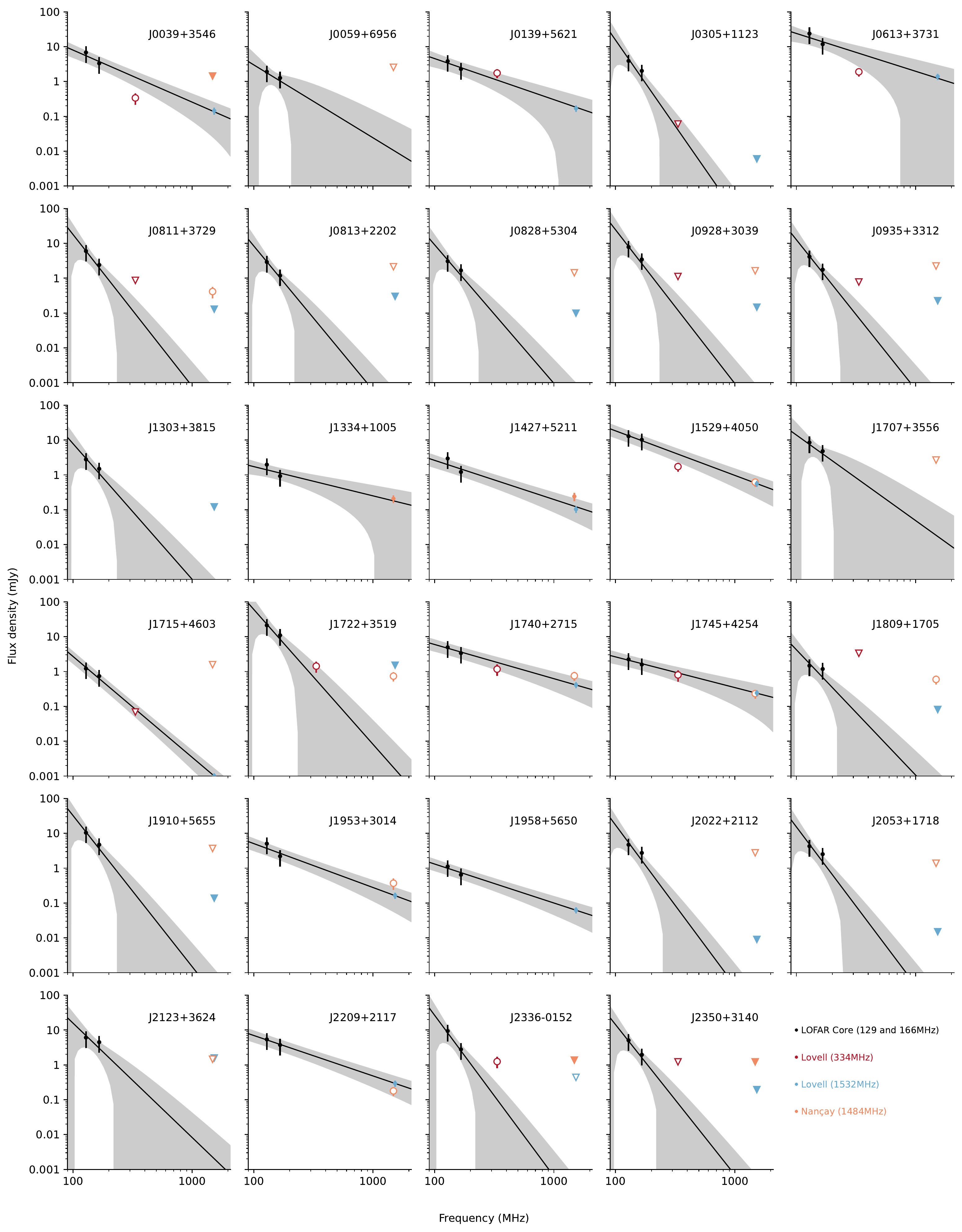}
  \caption{Mean flux densities (dots and circles) and upper limits for flux densities (triangles) as reported in Table\,\ref{tab:flux} for the 29 pulsars whose timing positions fall within the FWHM of the LOFAR Core tied-array beam. Filled dots and triangles are flux density measurements for which five or more observations were available, while empty circles and triangles are measurements from less than five observations. The black lines represent fitted spectral indices, which are the mean values for the spectral index $\alpha$ and the flux density on the reference frequency of 149\,MHz $S_0$ from the \textsc{emcee} samples.  The shaded regions represent the uncertainty on $\alpha$ and $S_0$ propagated to flux densities between 90 and 2100\,MHz. For the spectral index fit, only flux density measurements with five or more observations were used.}
    \label{fig:flux}
\end{figure*}

\subsection{Flux densities and spectral analysis}
\label{sec:ana_flux}
Mean flux densities were determined from the LOFAR Core, Lovell, and Nan\c cay observations. Observations from LOFAR were fully averaged in time and split in two subbands with centre frequencies of 129 and 166\,MHz. For each pulsar, all observations at these frequencies were combined and flux densities were calculated using the radiometer equation following the method from \citet{Kondratiev16}. This approach models the LOFAR beam, the system temperature, and the beamformer coherence, combined with sky temperatures taken from \citet{Price21}. We found that, on average, between 5 and 10\% of the dipoles in the LOFAR High Band Antennas were not in operation, reducing the effective area \citep{Kondratiev16}. We used a conservative estimate of 5\% of broken antennas for all flux density calculations as to not overestimate the flux densities of the pulsars. For observations during which 10\% of the dipoles were not operational, this would result in an underestimation of the flux density which is much lower than the uncertainty from other factors. We assumed a systematic uncertainty due to RFI, scintillation, and fluctuating tied array beam gain and system temperatures of 50\%, as suggested in \citet{Kondratiev16}. 

Note that we do not attempt to correct the LOFAR flux densities for positional offsets of the pulsars with respect to the centre of the LOFAR Core tied-array beam. Positional offsets are introduced by ionospheric effects, which can introduce random displacements of the order of $\sim1\arcmin$ compared to the $3\farcm5$ full width half maximum (FWHM) of the tied-array beam during observations, but more importantly, there are systematic offsets between the final timing position and the position determined from gridding and used to obtain the timing observations. For six pulsars, the gridding positions place them outside of the FWHM of the tied-array beam. While the corrections using a squared $\mathrm{sinc}$ function were used to correct flux densities of known pulsars in \citet{Sanidas19}, this approach is not suitable here, as the FWHM of a tied-array beam from the LOFAR Core stations is a factor 6 smaller than that from the stations on the LOFAR Superterp. Hence, we caution the reader to treat the LOFAR flux densities as indicative and that they should not be used for detailed studies. We expect that accurate flux densities of the LOTAAS pulsars will become available with the continued data releases from the LOFAR continuum imaging survey \citep{Shimwell17}.

For PSRs\,J0059+6956, J0305+1123, J0813+2202, J0828+5304, J0935+3312, J1427+5211, J1953+3014, J1958+5650, and J2123+3624, the observing position was updated part way through the observing time span, resulting in smaller positional offsets for the later observations. Only the later observations with improved gridding position were used to determine flux densities for these pulsars. 

From the Lovell observations, we obtained flux densities at 334 and 1532\,MHz. For each pulsar, the observations were fully averaged in time and frequency and combined using \textsc{psradd}. Flux densities were calculated from the radiometer equation using a gain of 0.8\,K\,Jy$^{-1}$ for both centre frequencies, an elevation-dependent system temperature to take into account spill-over, and sky temperatures from \citet{Price21}. The NRT observations were calibrated using a noise diode which injected white noise into the receiver to allow for the transfer of the flux scale using observations of flux calibrators. To estimate the uncertainty on the flux density measurements $\sigma_S$ in the Lovell and Nan\c cay observations, we applied a simplification of the method used in \citet{Jankowski18}: $\sigma_S = \sqrt{(0.2 S)^2 + (0.3S)^2/N}$, with $S$ the measured flux density for each pulsar, and $N$ the number of observations, which assumed a systematic uncertainty of 20\% and a typical modulation index due to scintillation of 30\%. In the case of non-detections in the Lovell and Nan\c cay observations, upper limits were determined by averaging all observations and calculating the $3\sigma$ limit on the root-mean-square noise of the pulse profile.

For those 29 pulsars whose timing positions fall within the FWHM of the LOFAR Core tied-array beam, we used Bayesian methods to model the radio spectra from both the flux density measurements as well as the upper limits. We used \textsc{emcee} \citep{ForemanMackey13} to maximise the likelihood function for detections and non-detections as described by \citet[][Eqn.\,1 to 3]{Laskar14}. We assumed that the radio spectra can be modelled with a single power law $S_\nu = S_0 \left(\nu /\nu_0 \right)^\alpha$,  where $S_\nu$ are the observed mean flux densities or upper limits at their respective observing frequencies $\nu$, with $\alpha$ as the spectral index for a flux density $S_0$ at reference frequency $\nu_0=149$\,MHz. Flat priors were assumed in $S_0$ and for $-9 < \alpha < 3$. We note that the simplification of a single power law may not be valid for all pulsars, as some pulsars may have spectra which are better described by a broken power law \citep[e.g.][]{Bilous16, Jankowski18}. 

\section{Results}
\label{sec:res}
\subsection{Timing}
\label{sec:res_timing}
The timing residuals shown in Fig.\,\ref{fig:residuals} using the timing models from Table\,\ref{tab:timing} and \ref{tab:J0828} are generally noise-like (i.e., uncorrelated and around the same level as the TOA uncertainties) for the majority of pulsars. The residuals of PSRs\,J0139+5621, J0613+3731, and J2306+3124 display time-dependent trends, while the timing residuals of PSRs\,J0811+3729, J0928+3039, J1722+3519, and J1910+5655 show non-Gaussian noise, resulting in timing solutions with high reduced $\chi^2$ values. 

In the case of PSR\,J0139+5621, the residuals of the LOFAR observations obtained after the observational gap between 2018 May and 2019 June show a downward trend, indicating the possibility of a glitch, a sudden increase in the spin frequency, which is often followed by a recovery towards the pre-glitch frequency \citep{Radhakrishnan69, Reichley69}. Fitting the spin-frequency $\nu$ with the TOAs before and after MJD 58257 separately, we obtained a fractional glitch size $\Delta \nu / \nu$ of $3.3\times10^{-10}$, which is near the lower values of the observed range of glitch sizes of $10^{-11}$ and $10^{-5}$ \citep{Espinoza11, Yu13, Basu22}. As glitches are predominantly seen in younger pulsars, and since PSR\,J0139+5621 is the youngest pulsar in our sample, with a characteristic age $\tau_c$ of $3.6\times10^5$\,yr (see below for further explanations on $\tau_c$), we consider it likely that a glitch occurred in PSR\,J0139+5621.

We considered the possibility that the long-term trends seen in the residuals for PSRs\,J0613+3731 and J2306+3124 could be due to magnetic dipole braking. This would yield a measurable second spin-frequency derivative $\ddot{\nu} = n \dot{\nu}^2/\nu$ for spin-frequency $\nu$ and spin-frequency derivative $\dot{\nu}$ with $n=3$. However fitting for $\ddot{\nu}$ in the timing solutions of PSRs\,J0613+3731 and J2306+3124 yields unrealistically large values for $|n|$, indicating that the observed trends are likely due to timing noise, which is seen in the majority of normal pulsars timed over long time spans \citep[e.g.,][]{Hobbs10}.

We found that the noise in the timing residuals of PSRs\,J0811+3729, J0928+3039, J1722+3519, and J1910+5655 is caused by nulling and pulse-to-pulse profile variations that do not average out over the length of the observation, see Fig.\,\ref{fig:YFp}. We discuss some of these pulsars in \S\,\ref{sec:profevol}.

For the majority of the pulsars the offset between the timing position and the gridded position used to obtain the timing observations (Table\,\ref{tab:flux}) fall within the $1\farcm75$ half-width at half maximum (HWHM) of the tied-array beams with the LOFAR Core. For PSR\,J0305+1123, the position used to obtain the observations was updated in 2017 November, resulting in higher signal-to-noise detections and more precise TOAs afterwards (Fig.\,\ref{fig:residuals}). Of the six outliers (PSRs\,J0317+1328, J1814+2224, J1958+2214, J2006+2205, J2057+2128, and J2306+3124), all have offsets below $2\farcm5$, except for PSR\,J1958+2214, which is offset by $16\farcm3$ and located in the far side-lobes of the LOFAR Core tied-array beam. The origin of this large offset was traced back to the LOTAAS survey observation in which PSR\,J1958+2214 was discovered, which showed several high signal-to-noise detections at widely separated tied-array beams. These odd side-lobe detections were noted at the time, and the choice was made to perform follow-up gridding observations using more narrow tied-array beams towards the detections of two neighbouring tied-array beams in the centre of the hexagonal grid of 61 LOFAR Superterp tied-array beams, while the actual position of PSR\,J1958+2214 turned out to be consistent with one of the detections in the outer ring of the hexagonal grid.

Significant measurements of proper motion were obtained for PSRs\,J0828+5304 and J2053+1718. The proper motion of PSR\,J0828+5304 is $\mu_\alpha \cos \delta=4.8\pm1.8$\,mas\,yr$^{-1}$ and $\mu_\delta=-14\pm3$\,mas\,yr$^{-1}$, yielding transverse velocities of 63 and 110\,km\,s$^{-1}$ for the 0.9 and 1.6\,kpc distances predicted by the NE2001 \citep{Cordes02} and YMW16 \citep{Yao17} Galactic electron density models. For PSR\,J2053+1718, with $\mu_\alpha \cos \delta=-31\pm5$\,mas\,yr$^{-1}$ and $\mu_\delta=15\pm9$\,mas\,yr$^{-1}$, the transverse velocities are 315 and 344\,km\,s$^{-1}$ at the NE2001 distance of 1.9\,kpc and the YMW16 distance of 2.1\,kpc. This pulsar, PSR\,J2053+1718, was first discovered by Arecibo \citep{Ray96} and a timing solution has recently been published by \citet{Brinkman18}. They obtained significantly lower transverse velocities of 63 and 69\,km\,s$^{-1}$ for the NE2001 and YMW16 distances, over a longer timing baseline. As \citet{Brinkman18} reports DM variations on the order of $5\times10^{-3}$\,pc\,cm$^{-3}$, we attribute the larger LOTAAS proper motion measurement, as well as the high reduced $\chi^2$ of the LOTAAS timing solution to unmodelled DM variations affecting the arrival time measurements. Based on the short spin period and low magnetic field, \citet{Brinkman18} propose that PSR\,J2053+1718 may be a disrupted recycled pulsar, a pulsar which was spun up in a binary system with a massive stellar companion which got disrupted and unbound when the companion went supernova. This origin could explain the low magnetic field of this system, which are typically seen in double neutron star systems (DNSs; see \citealt{Tauris17} for a review on DNSs). 

The period and period derivative from the timing models were used to calculate the characteristic age $\tau_\mathrm{c}$, the surface magnetic field strength $B$, and the intrinsic spin-down energy $\dot{E}$ (see the rightmost columns in Table \ref{tab:flux}. Here, we assumed a canonical neutron star with a radius of 10\,km and a moment of inertia of $10^{45}\,$ g\,cm$^2$. Furthermore, we assumed that the initial spin is negligible and that magnetic dipole radiation is responsible for the spin-down throughout the pulsar's lifetime (see \citealt{Lorimer12} for details). 

\subsection{Pulse profile variations}
\label{sec:profevol}
\begin{figure*}
  \centering
  \includegraphics[width=\textwidth]{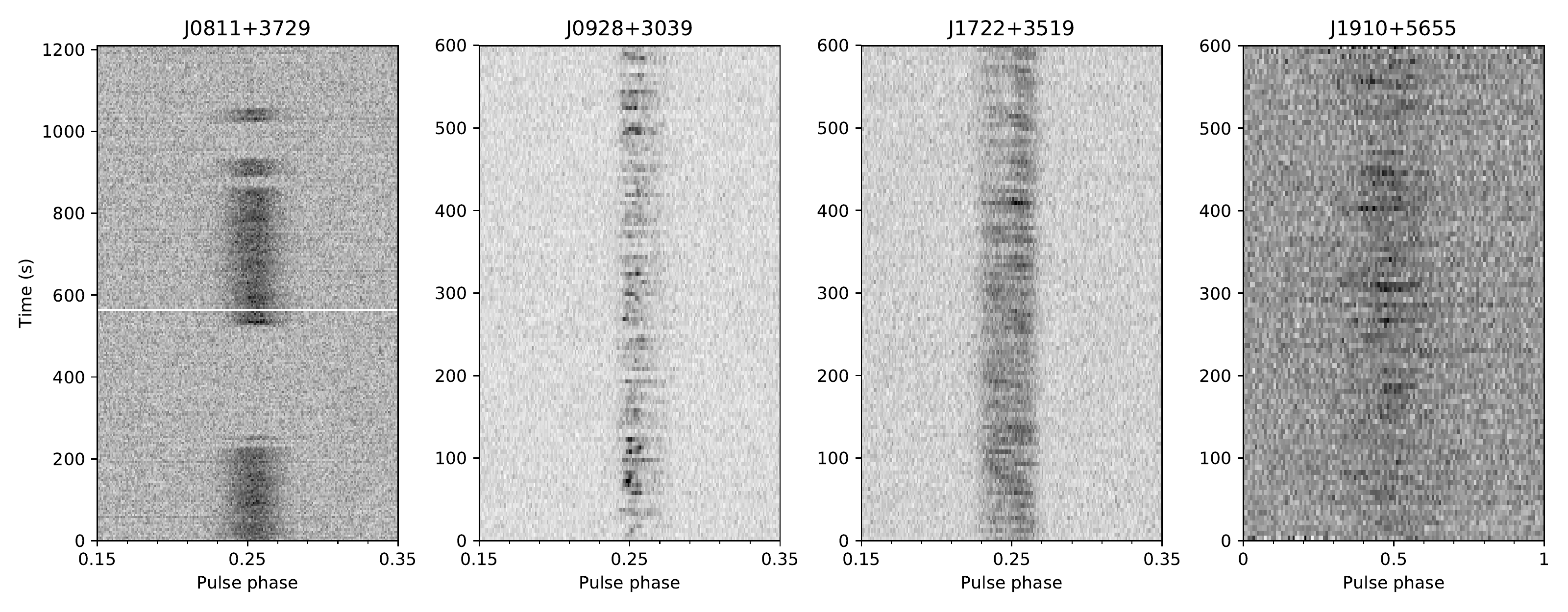}
  \caption{Pulse profile variations with time for four pulsars at 149\,MHz. The figures display multiple rotations per sub-integration. A portion of the full rotation is shown here and the profiles are rotated such that the highest peak is centred, at phase $\phi=0.25$ or $\phi=0.5$. The flux scale is normalised for each pulsar separately. White bars indicate where RFI was removed from the data.}
    \label{fig:YFp}
\end{figure*}

\begin{figure*}
  \centering
  \includegraphics[width=\textwidth]{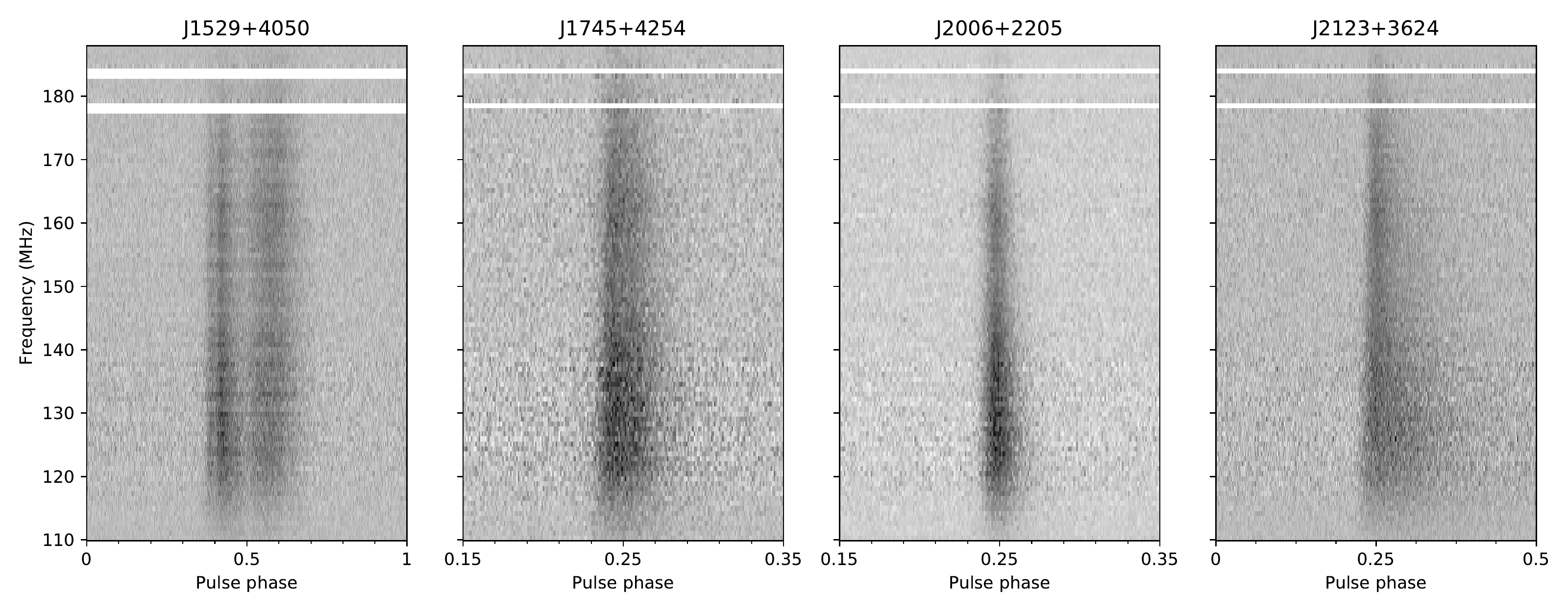}
  \caption{Pulse profile variations in frequency between 110 and 188\,MHz in four pulsars. A portion of the full rotation is shown here and the profiles are rotated such that the highest peak is centred, at phase $\phi=0.25$ or $\phi=0.5$. The flux scale is normalised for each pulsar separately. White bars indicate where RFI was removed from the data.}
    \label{fig:GTp}
\end{figure*}

The pulse profiles of some of the LOTAAS pulsars show profile variations in time or frequency. Figures\,\ref{fig:YFp} and \ref{fig:GTp} highlight these variations. 

PSR\,J0811+3729 is the most extreme case by displaying nulling, a relatively common phenomenon where the pulsed emission temporarily ceases, with nulls lasting between a single pulsar rotation up to several weeks (e.g \citealt{Kramer06, Young15, Basu17}). In the absence of observations with single pulse integrations, we analysed the energy distributions of the 5-s sub-integrations, where each sub-integration covers just over almost exactly 4 individual pulses. For each sub-integration, we determined if pulsed emission was present by applying the criteria from \citet{Wang20}, which compare the emission in the on-pulse region with the standard deviation of the emission in the off-pulse region. We found that for the total observation length of PSR\,J0811+3729 of 14.4\,h, 80.1\% of 5-s sub-integrations were in a null state. The mean length of a null state was 105\,s with the longest null lasting a full observation of 20\,min. The average time between nulls was approximately 29\,s and the longest time the pulsar was in an emission state was 5.67\,min.

For PSR\,J0928+3039, the 5-s sub-integrations covered 3.43 individual pulses and, as shown in a representative observation in Fig.\,\ref{fig:YFp}, these revealed strong variability on time-scales of tens of seconds. The strong variations appear to occur primarily in the leading half of the profile. Pulsed emission was also absent for occasional sub-integrations, indicating that this pulsar may be nulling for a few rotations at a time. No evidence for drifting sub-pulses was visible, but would require observations of individual pulses to completely rule out.

The profile of PSR\,J1722+3519 at 149\,MHz shows two narrow components whose amplitude varies between observations, resulting in the additional noise seen in the timing residuals. In most observations the leading component of the averaged pulse profile has half the amplitude of the trailing component, but in some observations both components have comparable amplitudes. The 5-s sub-integrations (6.08 individual pulses) reveal that these variations in the averaged profiles are caused by pulse-to-pulse variations between the individual pulse components. To understand the profile variability of PSR\,J1722+3519, a 1\,hr observation where individual pulses were recorded was obtained and analysed. This reveals a highly variable intensity of the two profile peaks with a relatively long timescale of $\sim100$ pulse periods. A fluctuation spectrum analysis using the methods described in \citet{Weltevrede16} confirmed the presence of this modulation timescale. For many pulsars, periodic intensity modulation can be associated with drifting sub-pulses (e.g. Song et al., submitted). However, for PSR\,J1722+3519, only evidence for amplitude modulation is found. Given the long timescale of the modulation, even longer observations would be beneficial in order to reveal any associated phase variations. Another possible explanation for the profile variation and the large root-mean-square (RMS) residuals is short timescale mode switching, as seen in for example \citet{Stairs19}.  

In the case of PSR\,J1910+5655, 14.63 individual pulses were averaged into the 5-s sub-integrations, which still show significant variability between sub-integrations. The relatively low signal-to-noise of the profile in the 5-s sub-integrations, combined with the averaging of many pulses per sub-integration and the single wide component (21\% duty cycle) make it difficult to classify these variations. 

The pulse profile of PSR\,J1529+4050 displays multiple components at 149\,MHz, and in the LOFAR band between 110 and 188\,MHz, the leading component decreases in brightness with increasing frequency compared to the trailing component (Fig.\,\ref{fig:GTp}). At 334\,MHz and 1532\,MHz, only the trailing component remains (Fig.\,\ref{fig:profiles}), indicating that the leading component has significantly steeper spectra compared to the trailing component. In a study comparing the pulse profiles of a hundred pulsars observed by the LOFAR HBAs as well as the Lovell telescope at L-band, \citet{Pilia16} found pulse profiles with more profile components at HBA frequencies compared to L-band frequencies to be less common than the other way around. More components at HBA frequencies may be explained by a wider pulsar beam at lower frequencies \citep{Cordes78}, resulting in more components coming into view. \citet{Pilia16} note a slight indication that pulsars with more components at HBA frequencies than at L-band frequencies may be older and have a longer period. With a characteristic age of $2.8\times10^{9}$\,yr, PSR\,J1529+4050 is the fourth oldest pulsar in our sample.

\subsection{Pulse widths}
\label{sec:scattering}
\begin{figure}
  \centering
  \includegraphics[width=\columnwidth]{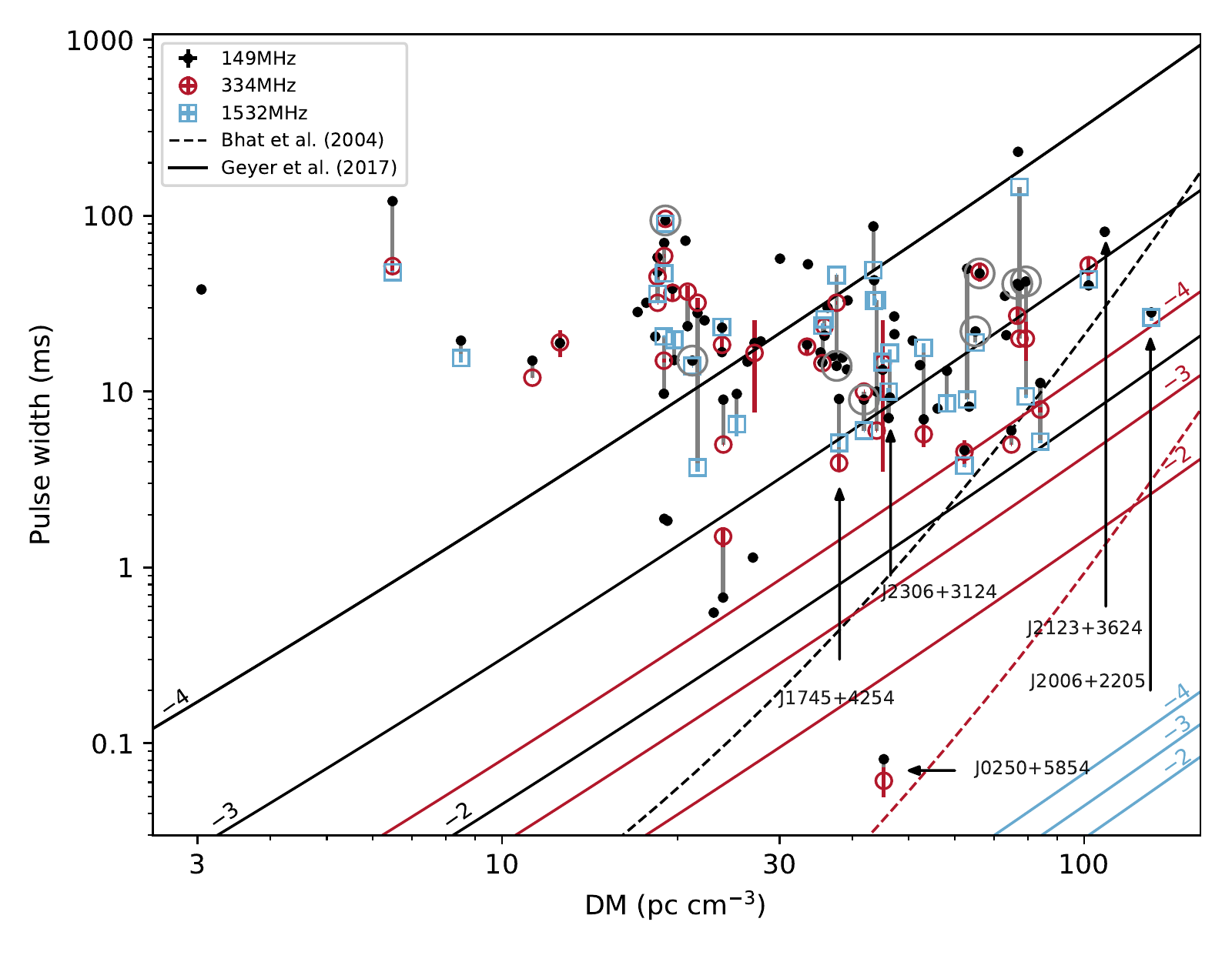}
  \caption{The relations between pulse broadening and DM as found by \citet{Bhat04} (dashed lines) and \citet{Geyer17} (solid lines) at 149, 334, and 1532\,MHz in black, red, and blue respectively. The functions from \citet{Geyer17} are shown with scattering indices between $-4$ and $-2$, while the function from \citet{Bhat04} is displayed with the best-fitting index of $-3.68$. Shown as dots, circles and squares are the $W_{50}$ pulse widths from this study, as well as from the previous LOTAAS reports \citet{Tan18}, \citet{Michilli20}, and \citet{Tan20} at different frequencies. These measurements represent upper limits on pulse broadening due to interstellar scattering. Pulse widths from the same pulsar at different frequencies are connected with grey, vertical lines. The pulsars discussed in \S\,\ref{sec:scattering} are marked with arrows. The pulse widths at 149\,MHz of eight pulsars from \citet{Michilli20} and \citet{Tan20} which may be affected by scattering are marked with grey circles.}
    \label{fig:scattering}
\end{figure}

Figure\,\ref{fig:scattering} shows the pulse widths at 50\% of the peak profile intensity of all LOTAAS discoveries, combining the $W_{50}$ values from this paper with those reported in \citet{Michilli20}, \citet{Tan18}, and \citet{Tan20}. Because the $W_{50}$ of PSRs\,J0250+5854 and J1658+3630 were reported at slightly different observing frequencies, we used the $W_{50}$ value at a frequency nearest to the frequencies reported here. We found that, on average, the pulse widths are somewhat wider at 149\,MHz than at the higher frequencies, with a mean $W_{50}$ at 149\,MHz of 29\,ms and a standard deviation of 33\,ms, compared to $24\pm20$\,ms and $27\pm28$\,ms at 334 and 1532\,MHz, respectively. 

The intrinsic profile of pulsars will be broadened by the effects of interstellar scattering, resulting in exponential tails at the trailing edge of the pulse profiles. Four pulsars from the sample reported here, PSRs\,J1745+4254, J2006+2205, J2123+3624, and J2306+3124, show these effects in the pulse profiles at 149, 334, and/or 1484 or 1532\,MHz (see Fig.\,\ref{fig:profiles} and Fig.\,\ref{fig:GTp}), suggesting that these may be scatter broadened. These pulsars are indicated in Fig.\,\ref{fig:scattering}, as are eight pulsars that \citet{Michilli20} and \citet{Tan20} reported as showing signs of scatter broadening. All these pulsars are at DMs above 20\,pc\,cm$^{-3}$, as expected, since scatter broadening is caused by material in the interstellar medium and hence is roughly correlated with the dispersion measure (DM).

Figure\,\ref{fig:scattering} also displays the scatter broadening predictions at 149, 334, and 1532\,MHz from the empirical relations between scattering and DM found by \citet{Bhat04} and \citet{Geyer17}. Most pulse widths displayed in Fig.\,\ref{fig:scattering} are around the same order of magnitude as the scatter broadening predicted by \citet{Bhat04} and \citet{Geyer17} for the typical scattering indices of $\alpha_\mathrm{S}\sim-4$ or $-4.4$ (for scattering timescales $\tau_\nu\propto\nu^{\alpha_\mathrm{S}}$) from the thin-screen approximation \citep{Williamson72}. Hence, the scattering relations predict that the pulse profiles of the majority of the pulsars in our sample should be affected by interstellar scattering. However, most pulse profiles in Fig.\,\ref{fig:profiles}, as well as the pulse profiles from the previous LOTAAS timing papers, do not show signs of scattering, suggesting that there is a less steep frequency scaling ($\alpha_\mathrm{S}>-4$) or that the scattering timescales may be smaller than predicted by \citet{Bhat04} and \citet{Geyer17}.

To test this, we modelled the frequency dependence of the LOFAR pulse profiles of the pulsars that show exponentially trailing edges, PSRs\,J1745+4254, J2006+2205, J2123+3624, and J2306+3124, with the wideband pulsar timing code \textsc{PulsePortraiture} by \citet{Pennucci19}, which simultaneously measures TOAs, DM, and scattering, by fitting frequency dependent templates. With this method, we constrained a scattering time $\tau_\mathrm{S}$ and scattering index $\alpha_\mathrm{S}$ for these four pulsars by modelling the intrinsic profile as a single Gaussian function and the pulse broadening as a one-sided exponential function. We fitted this model to an averaged pulse profile combining all LOFAR observations, which were fully averaged in time and partially averaged in frequency to 100 frequency channels. 

We found that the LOFAR profile of PSR\,J1745+4254 is best modelled with a large scattering index of $\alpha_\mathrm{S}=-1.10\pm0.12$ and a scattering timescale of $\tau_\mathrm{S}=5.49\pm8$\,ms. Similarly, PSR\,J2006+2205, has a large scattering index of $\alpha_\mathrm{S} = -2.18\pm15$ for $\tau_\mathrm{S} = 15.8\pm3$\,ms, while PSR\,J2123+3624 has a more typical scattering index of $\alpha_\mathrm{S}=-3.55\pm9$ for $\tau_\mathrm{S}=89.6\pm1.0$\,ms. We note that the profile of PSR\,J2306+3124 displays a similar exponential tail at both 149\,MHz and 1532\,MHz, with no discernible frequency evolution between the observing bands, indicating its intrinsic profile may have an exponential tail. The large scattering indices of PSRs\,J1745+4254 and J2006+2205 confirm that the effect of scatter broadening in the LOFAR pulse profiles is less pronounced compared to predictions from the typical $-4$ and $-4.4$ scattering index values, behaviour also reported in other pulsars at low observing frequencies \citep{Geyer17}.

\subsection{Spectral index}
Table\,\ref{tab:flux} displays the flux densities and the spectral indices and Fig.\,\ref{fig:flux} shows the mean flux densities and the spectral power laws. For the 29 pulsars investigated, here, we obtained a mean spectral index of $\alpha=-2.8\pm1.5$, with the uncertainty representing the standard deviation, which is consistent with the values of $\alpha=-2.4\pm0.9$ and $\alpha=-1.9\pm0.5$ from \citet{Michilli20} and \citet{Tan20}, respectively, despite slightly different treatments of the flux calibration and spectral fitting. Of the pulsars with spectral index measurements, PSR\,J2336$-$0152 displays a notably steep spectrum with a spectral index of $-5\pm2$, due to the large difference between the measured flux densities at 129 and 166\,MHz. However, these flux densities  have large uncertainties and a much flatter spectrum is also consistent with our results. We caution that a simple power law may be insufficient to describe the spectra of some pulsars.

\begin{table*}
   \footnotesize
        \caption{The results from the spectral analysis and parameters derived from the timing model in Table\,\ref{tab:timing}. Mean flux densities and upper limits on the mean flux densities $S_\nu$ are shown with observing frequencies in MHz indicated in subscript. The inferred spectral indices $\alpha$ were fitted with a single power law of the form $S_\nu\propto\nu^\alpha$. Flux densities for which less than five observations were available are indicated in italics and were not used in the spectral fit. The Offset in arcminutes refers to the difference between the centre of the telescope beam and the position reported in Table\,\ref{tab:timing}. For six pulsars, the spectral index was not calculated due to the large position offset. We determined the characteristic age $\tau_\mathrm{c}$, the surface magnetic field strength $B$, and the intrinsic spin-down energy $\dot{E}$ assuming a canonical neutron star with a radius of 10\,km and a moment of inertia of $10^{45}\,$ g\,cm$^2$. Furthermore, we assumed that the initial spin is negligible and that the spin-down is due to magnetic dipole radiation throughout the pulsar's lifetime. See \citet{Lorimer12} for further details.}
    \begin{tabular}{llllllll|lll}
        \hline
        PSR & S$_{129}$ &S$_{166}$& S$_{334}$& S$_{1484}$& S$_{1532}$&$\alpha$ & Offset & $\mathrm{\tau_c} $ & $\mathrm{B}$ &$ \mathrm{\dot{E}}$  \\
       & mJy & mJy & mJy& mJy & mJy & & arcmin  & (yr) & (G) & (erg s$^{-1})$ \\
        \hline
J0039+3546&$7(3)$&$3.3(1.7)$&$\it{0.34(12)}$&$<1.4$&$0.14(3)$&$-1.5(3)$&$1.12$&$1.3\times10^{8}$&$1.9\times10^{11}$&$1.6\times10^{31}$\\
J0059+6956&$1.9(1.0)$&$1.3(6)$&$-$&$<\it{2.5}$&$-$&$-2(3)$&$1.61\times10^{-3}$&$1.3\times10^{8}$&$4.1\times10^{11}$&$3.7\times10^{30}$\\
J0139+5621&$3.8(1.9)$&$2.3(1.1)$&$\it{1.8(5)}$&$-$&$0.17(3)$&$-1.2(2)$&$7.7\times10^{-1}$&$3.6\times10^{5}$&$1.2\times10^{13}$&$5.6\times10^{32}$\\
J0305+1123&$3.9(2.0)$&$2.0(1.0)$&$<\it{0.06}$&$-$&$<0.006$&$-4.8(1.8)$&$5.7\times10^{-1}$&$4.5\times10^{7}$&$5.2\times10^{11}$&$1.9\times10^{31}$\\
J0317+1328&$2.0(1.0)$&$1.1(5)$&$\it{0.9(3)}$&$<\it{2.7}$&$<\it{0.21}$&$-$&$1.84$&$1.8\times10^{8}$&$5.9\times10^{11}$&$8.7\times10^{29}$\\
J0613+3731&$24(12)$&$12(6)$&$\it{1.9(5)}$&$-$&$1.4(3)$&$-1.0(2)$&$4.3\times10^{-2}$&$3.0\times10^{6}$&$1.4\times10^{12}$&$5.4\times10^{32}$\\
J0811+3729&$6(3)$&$2.4(1.2)$&$<\it{0.86}$&$\it{0.41(15)}$&$<0.13$&$-4.4(1.9)$&$1.16$&$2.7\times10^{7}$&$9.8\times10^{11}$&$1.5\times10^{31}$\\
J0813+2202&$2.9(1.5)$&$1.2(6)$&$-$&$<\it{2.1}$&$<0.3$&$-4(2)$&$2.2\times10^{-1}$&$1.1\times10^{8}$&$2.1\times10^{11}$&$2.0\times10^{31}$\\
J0828+5304&$3.1(1.5)$&$1.7(8)$&$-$&$<\it{1.4}$&$<0.099$&$-4.0(1.9)$&$2.4\times10^{-4}$&$8.3\times10^{9}$&$6.0\times10^{8}$&$4.1\times10^{32}$\\
J0928+3039&$8(4)$&$3.4(1.7)$&$<\it{1.1}$&$<\it{1.6}$&$<0.15$&$-4.3(1.9)$&$8.5\times10^{-1}$&$4.6\times10^{7}$&$6.2\times10^{11}$&$1.3\times10^{31}$\\
J0935+3312&$4(2)$&$1.8(9)$&$<\it{0.77}$&$<\it{2.2}$&$<0.22$&$-4.2(2.0)$&$1.01$&$4.0\times10^{7}$&$6.1\times10^{11}$&$1.7\times10^{31}$\\
J1303+3815&$2.8(1.4)$&$1.5(7)$&$-$&$-$&$<0.12$&$-3.9(1.9)$&$1.25$&$1.8\times10^{7}$&$3.8\times10^{11}$&$2.2\times10^{32}$\\
J1334+1005&$2.0(1.0)$&$0.9(5)$&$-$&$0.21(5)$&$-$&$-0.9(6)$&$5.8\times10^{-1}$&$1.9\times10^{7}$&$8.3\times10^{11}$&$3.9\times10^{31}$\\
J1427+5211&$3.0(1.5)$&$1.2(6)$&$-$&$0.24(6)$&$0.10(2)$&$-1.1(3)$&$4.8\times10^{-3}$&$3.3\times10^{7}$&$7.0\times10^{11}$&$1.9\times10^{31}$\\
J1529+4050&$13(6)$&$10(5)$&$\it{1.7(5)}$&$\it{0.62(16)}$&$0.56(11)$&$-1.3(2)$&$1.29$&$2.8\times10^{9}$&$3.6\times10^{10}$&$9.7\times10^{29}$\\
J1707+3556&$8(4)$&$5(2)$&$-$&$<\it{2.7}$&$-$&$-2(3)$&$5.3\times10^{-1}$&$3.0\times10^{9}$&$1.2\times10^{10}$&$8.3\times10^{30}$\\
J1715+4603&$1.2(6)$&$0.7(4)$&$<\it{0.069}$&$<\it{1.6}$&$0.0010(2)$&$-2.9(3)$&$9.7\times10^{-1}$&$1.8\times10^{8}$&$1.6\times10^{11}$&$1.2\times10^{31}$\\
J1722+3519&$21(11)$&$11(5)$&$\it{1.4(5)}$&$\it{0.7(2)}$&$<1.5$&$-4(2)$&$5.5\times10^{-1}$&$2.4\times10^{7}$&$6.8\times10^{11}$&$3.9\times10^{31}$\\
J1740+2715&$5(2)$&$3.4(1.7)$&$\it{1.2(4)}$&$\it{0.8(2)}$&$0.41(9)$&$-0.99(20)$&$1.57$&$7.8\times10^{7}$&$4.8\times10^{11}$&$7.1\times10^{30}$\\
J1745+4254&$2.2(1.1)$&$1.6(8)$&$\it{0.8(3)}$&$\it{0.23(7)}$&$0.24(5)$&$-0.9(2)$&$1.5$&$4.9\times10^{8}$&$5.5\times10^{10}$&$1.4\times10^{31}$\\
J1809+1705&$1.5(7)$&$1.2(6)$&$<\it{3.3}$&$\it{0.59(17)}$&$<0.081$&$-3.5(1.9)$&$1.74$&$1.1\times10^{7}$&$2.6\times10^{12}$&$1.4\times10^{31}$\\
J1814+2224&$1.1(5)$&$1.3(7)$&$\it{0.8(3)}$&$\it{0.28(10)}$&$0.20(4)$&$-$&$2.4$&$1.1\times10^{7}$&$3.0\times10^{11}$&$8.6\times10^{32}$\\
J1910+5655&$10(5)$&$5(2)$&$-$&$<\it{3.7}$&$<0.14$&$-4.3(1.8)$&$1.22$&$2.6\times10^{7}$&$2.7\times10^{11}$&$2.0\times10^{32}$\\
J1953+3014&$5(3)$&$2.2(1.1)$&$-$&$\it{0.37(13)}$&$0.16(3)$&$-1.3(3)$&$1.92\times10^{-3}$&$5.2\times10^{7}$&$7.1\times10^{11}$&$7.4\times10^{30}$\\
J1958+2214&$0.6(3)$&$0.8(4)$&$-$&$-$&$-$&$-$&$1.63\times10^{1}$&$3.6\times10^{6}$&$2.2\times10^{12}$&$1.6\times10^{32}$\\
J1958+5650&$1.1(6)$&$0.6(3)$&$-$&$-$&$0.063(13)$&$-1.1(2)$&$3.0\times10^{-3}$&$2.9\times10^{7}$&$2.3\times10^{11}$&$2.2\times10^{32}$\\
J2006+2205&$3.7(1.9)$&$3.0(1.5)$&$<\it{2.3}$&$0.19(7)$&$<0.12$&$-$&$2.0$&$4.8\times10^{6}$&$3.2\times10^{12}$&$4.3\times10^{31}$\\
J2022+2112&$5(2)$&$2.7(1.4)$&$-$&$<\it{2.8}$&$<0.009$&$-4.6(1.7)$&$8.8\times10^{-1}$&$1.4\times10^{7}$&$8.6\times10^{11}$&$6.9\times10^{31}$\\
J2053+1718&$4(2)$&$2.5(1.3)$&$-$&$<\it{1.4}$&$<0.015$&$-4.4(1.7)$&$7.4\times10^{-1}$&$6.3\times10^{9}$&$6.1\times10^{9}$&$7.0\times10^{30}$\\
J2057+2128&$1.6(8)$&$0.8(4)$&$-$&$<3.0$&$-$&$-$&$2.3$&$3.5\times10^{7}$&$7.9\times10^{11}$&$1.3\times10^{31}$\\
J2123+3624&$6(3)$&$4(2)$&$-$&$<\it{1.4}$&$<1.6$&$-3(2)$&$8.2\times10^{-3}$&$9.8\times10^{6}$&$1.7\times10^{12}$&$3.8\times10^{31}$\\
J2209+2117&$5(3)$&$3.7(1.9)$&$-$&$\it{0.18(5)}$&$0.29(6)$&$-1.17(19)$&$6.8\times10^{-1}$&$5.5\times10^{6}$&$3.0\times10^{12}$&$3.6\times10^{31}$\\
J2306+3124&$6(3)$&$3.3(1.7)$&$-$&$\it{0.18(7)}$&$0.18(4)$&$-$&$2.0$&$2.4\times10^{7}$&$2.8\times10^{11}$&$2.2\times10^{32}$\\
J2336$-$0151&$9(5)$&$2.8(1.4)$&$\it{1.2(4)}$&$<1.4$&$<\it{0.44}$&$-5(2)$&$1.54$&$4.7\times10^{7}$&$6.0\times10^{11}$&$1.2\times10^{31}$\\
J2350+3140&$5(3)$&$1.9(1.0)$&$<\it{1.2}$&$<1.2$&$<0.2$&$-4.2(2.0)$&$1.01$&$2.3\times10^{7}$&$4.3\times10^{11}$&$1.1\times10^{32}$\\
        \hline
    \end{tabular}
    \label{tab:flux}
\end{table*}

\subsection{PSR\,J0828+5304}
PSR\,J0828+5304 is a binary millisecond pulsar with a spin period of 13.52\,ms at a DM of 23.103\,pc\,cm$^{-3}$. Table\,\ref{tab:J0828} provides the binary parameters of this pulsar. The orbital period of $P_\mathrm{b}=5.899$\,d and projected semi-major axis of $x=13.147$\,lt-s yield a mass-function of 0.07\,M$_\odot$, constraining the binary companion mass to 0.65\,M$_\odot$ for an edge-on orbit, or 0.79\,M$_\odot$ for a median orbital inclination of 60$\degr$, assuming the pulsar has a canonical mass of 1.4\,M$_\odot$. With this companion mass and spin period, PSR\,J0828+5304 can be classified as an intermediate mass binary pulsar \citep{vanKerkwijk05} where the neutron star is partially recycled due to unstable mass transfer from a binary companion which will leave a CO or ONeMg-core white dwarf \citep{Tauris12}. The Laplace-Lagrange parameters of the ELL1 binary model for nearly circular orbits yield an orbital eccentricity of $2.5(7)\times10^{-6}$, in line with predictions from tidal orbital circularisation \citep{Phinney92,Phinney94}.

No optical counterpart is present at the pulsar timing position of PSR\,J0828+5304 in the Pan-STARSS1 survey (PS1; \citealt{Chambers16}); the nearest catalogued PS1 source is $24\arcsec$ offset. The nominal PS1 $5\sigma$ detection limit of r-band magnitude $r=23.2$ provides absolute magnitude limits of $M_r>12.2$ for the 0.9 and 1.6\,kpc distances estimated from the observed DM and the NE2001 \citep{Cordes02} and YMW16 \citep{Yao17} Galactic electron density models, respectively. At this limit, cooling models\footnote{\url{http://www.astro.umontreal.ca/\~bergeron/CoolingModels}} by \citet{Bergeron11} and \citet{Tremblay11} for 0.7\,M$_\odot$ white dwarfs with either hydrogen or helium atmospheres predict effective temperatures and white dwarf cooling ages of $T_\mathrm{eff}<12000$\,K and $\tau_\mathrm{cool}>0.5$\,Gyr, ruling out young and hot white dwarf companions.

\section{Discussion and conclusions}
\label{sec:disc}
We present the properties of 35 pulsars discovered with LOFAR as part of the LOTAAS \citep{Sanidas19} and LOFAR pilot surveys \citep{Coenen14}. We expand this sample with the properties of 41 LOFAR-discovered pulsars reported by \citet{Michilli20}, \citet{Tan18}, and \citet{Tan20}, all of which were discovered by the LOTAAS survey. We excluded PSR\,J0815+4611, whose timing parameters were presented in \cite{Michilli20}, as it was discovered as a steep spectrum polarised point source in continuum images of the 3C\,196 field \citep{Jelic15}. Similarly, we excluded radio pulsars discovered with LOFAR and presented by \citet{Pleunis17}, \citet{Bassa17,Bassa18} and \citet{Sobey22} as these pulsars were discovered through the use of coherent dedispersion and by targeting the locations of Fermi $\gamma$-ray sources or polarised point sources from the LOFAR imaging survey \citep{Shimwell17} and will have different survey selection effects compared to LOTAAS.

\begin{figure}
    \centering
    \includegraphics[width=\columnwidth]{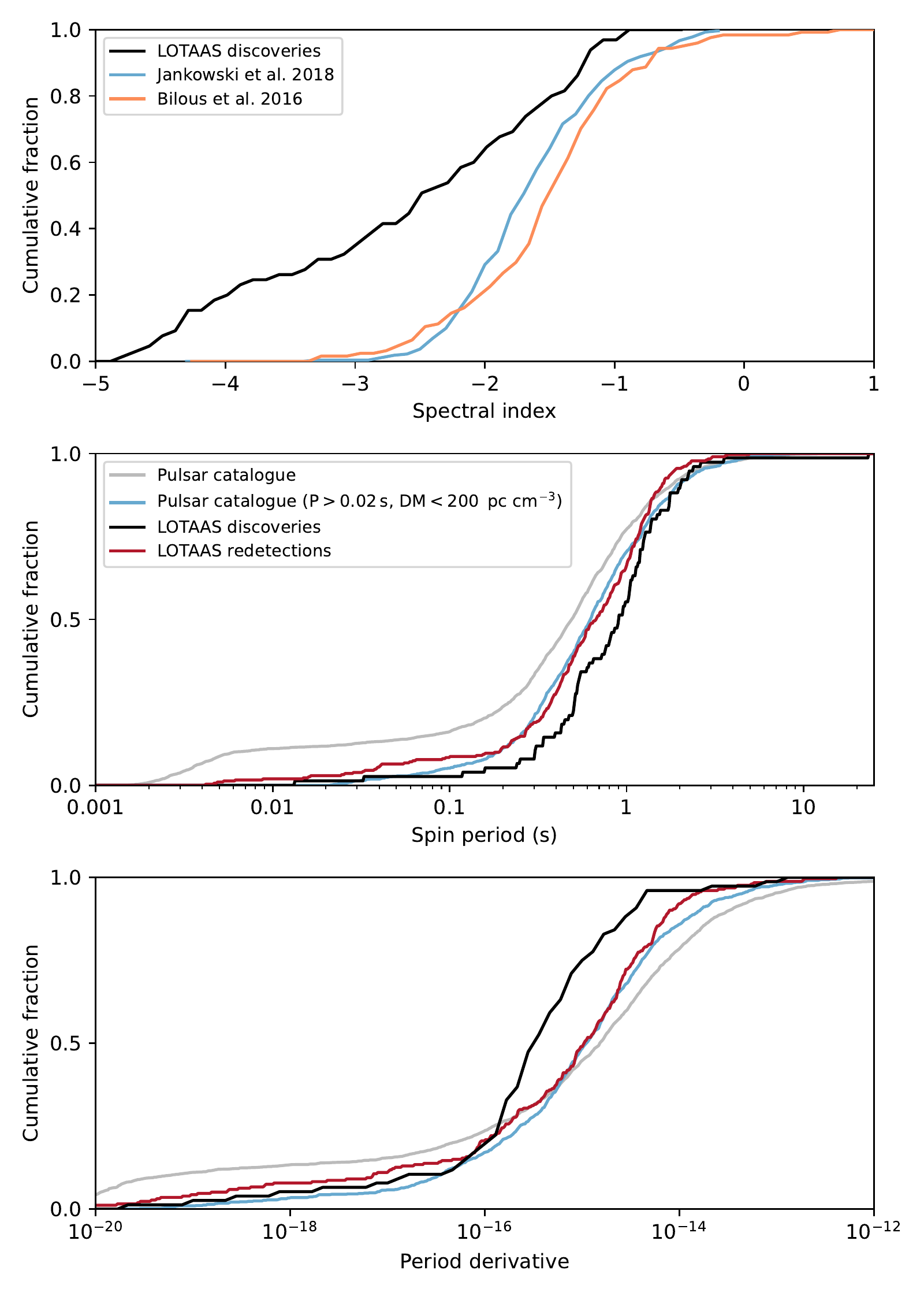}
    \caption{Cumulative distributions of the spectral index, the spin period, and the period derivative of the pulsars discovered by the LOTAAS survey (black). The upper plot displays spectral indices from \citet{Jankowski18} (blue) and \citet{Bilous16} (orange). We caution that, due to the uncertainties in the flux calibration of the LOTAAS obervations, the LOTAAS spectral indices should not be used for detailed studies. The lower plots show the LOTAAS redetections (red), the known pulsar population (grey), and a subset of the known pulsar population limited to $P>0.02$\,s and $\mathrm{DM}<200\,$pc\,cm$^{-3}$ (blue).}
    \label{fig:ppdot_dis}
\end{figure}

The spectral properties of the LOTAAS sample have a mean spectral index, assuming power law spectra of the form $S_\nu\propto\nu^\alpha$, of $\alpha=-2.5$ with a standard deviation of 1.2 (for 66 pulsars), see Fig.\,\ref{fig:ppdot_dis}. On average, these spectra are steeper compared to the sample studied by \citet{Jankowski18}, which has $\alpha=-1.57\pm0.62$ for 441 radio pulsars at radio frequencies between 728 and 3100\,MHz, as well as those of single power law spectral fits by \citet{Bilous16}, which have $\alpha=-1.4\pm0.7$ for 124 pulsars, all with observations from the LOFAR band between 110 and 188\,MHz, the same band used for the LOTAAS timing observations presented here. 

The steeper radio spectra of LOTAAS-discovered pulsars are readily explained as a selection effect, as pulsars with flatter spectra will already have been discovered by earlier pulsar surveys at higher frequencies. In particular, the LOTAAS survey overlaps with the GBNCC survey of the Northern sky using the Green Bank telescope \citep{Stovall14}, which started in 2009 and reaches a sensitivity of 1.1\,mJy at 350\,MHz, the AO327 survey with Arecibo, reaching 0.4\,mJy at 327\,MHz for declinations between $-1\degr$ and $28\degr$, which started survey observations in 2011 \citep{Deneva13}, and PUMPS with the Large Phased Array at Pushchino, which started in 2014 at 111\,MHz with sensitivities up to 0.1\,mJy for declinations between $-9\degr$ and $42\degr$ \citep{Tyul'bashev22}. Of the 311 known radio pulsars redetected with LOTAAS, 47 and 6 were earlier discoveries from the GBNCC and AO327 surveys, respectively \citep{Sanidas19}, indicating that these pulsars would have been new discoveries for LOTAAS in the absence of the GBNCC and AO327 surveys. Similarly, those pulsars discovered with LOTAAS and detected at 334\,MHz with the Lovell telescope have 334\,MHz flux densities near the sensitivity limits of the GBNCC and AO327 surveys, indicating they likely were below their detection thresholds.

The timing solutions of the LOTAAS pulsars show that they have, on average, longer spin periods and lower spin period derivatives in comparison to the known population of radio pulsars. These findings confirm the results from \citet{Sanidas19}, who already noted the statistical discrepancy of the LOTAAS discovery spin periods being longer than those from the known population. To compare the LOTAAS spin periods and spin period derivatives with those of the known pulsar population, we constructed three pulsar samples; \textsc{I}) the LOTAAS discoveries; \textsc{II}) known pulsars that were redetected in the LOTAAS survey and \textsc{III}) known radio pulsars that were not detected in LOTAAS, and that have $P>0.02$\,s and $\mathrm{DM}<200$\,pc\,cm$^{-3}$, to take into account some of the selection biases of LOTAAS, namely the increased sensitivity towards slower pulsars and decreased sensitivity for pulsars at high DM \citep{Sanidas19}. We queried the ATNF pulsar catalogue\footnote{\url{http://www.atnf.csiro.au/research/pulsar/psrcat}} (version 1.67, \citealt{Manchester05}) for the properties of the known pulsars to construct samples \textsc{II} and \textsc{III}, where we excluded pulsars that were associated with globular clusters. For sample \textsc{II}, the redetected pulsars presented in \citet{Sanidas19}, spin period and spin period derivatives were available for 253 out of the 311 pulsars, while sample \textsc{III} consists of 891 pulsars.

Figure\,\ref{fig:ppdot_dis} shows the cumulative distributions of the spin period and spin period derivatives of the three samples, as well as the entire known pulsar population. Using two-sample Kolmogorov-Smirnov (KS) tests, we found that the probability that the spin periods and spin period derivatives of sample \textsc{I}, the LOTAAS discoveries, are drawn from the selection-bias-corrected known pulsar population (sample \textsc{III}) is low, at $p=0.73$\% and 0.039\%, respectively. As these probabilities are $p<1$\%, we can reject the null hypothesis that the samples were drawn from the same underlying distribution. For comparison, the probability that the known pulsars redetected by LOTAAS, sample \textsc{II}, are drawn from sample \textsc{III} is significantly higher, $p=33$\% for the spin period distribution and 30\% for the spin period derivative distribution. 

Assuming that the spin period $P$ and spin period derivative $\dot{P}$ distributions from Fig.\,\ref{fig:ppdot_dis} both follow a log-normal distribution, we found $\log_{10} P=-0.040\pm0.421$ and $\log_{10} \dot{P}=-15.43\pm1.10$ for the median spin period ($P$ in s) and spin period derivatives ($\dot{P}$ in s\,s$^{-1}$) of the LOTAAS discoveries (sample \textsc{I}), compared to $\log_{10} P=-0.20\pm0.43$ and $\log_{10} \dot{P}=-14.94\pm1.18$ for the known pulsar population with $P>0.02$\,s and $\mathrm{DM}<200$\,pc\,cm$^{-3}$ (sample \textsc{III}). Hence, the LOTAAS-discovered pulsars have spin periods and spin-down rates which are, on average, a factor 1.4 longer and a factor 3.1 lower, respectively, compared to the known pulsar population.

\begin{figure}
  \centering
  \includegraphics[width=\columnwidth]{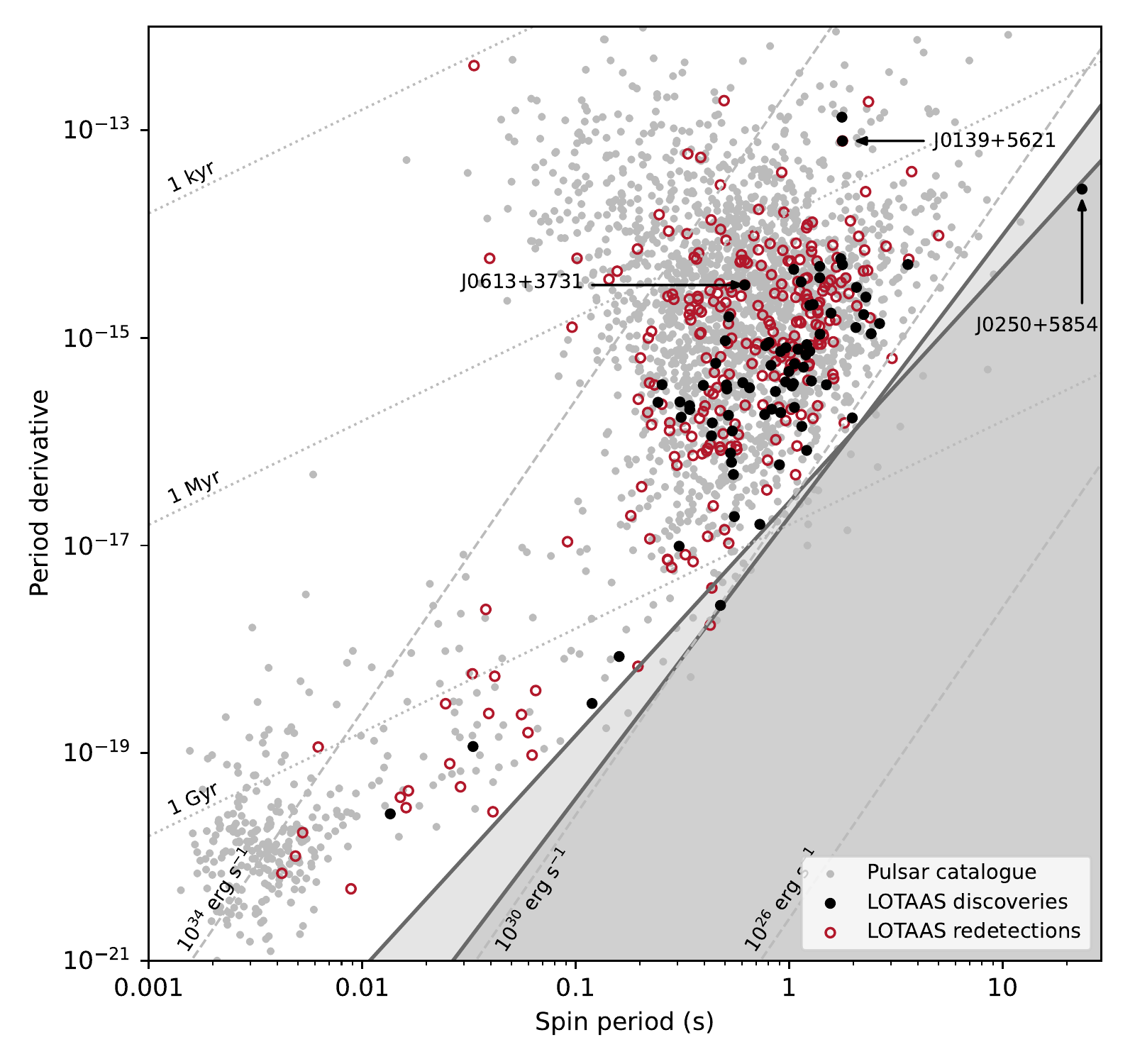}
  \caption{$P-\dot{P}$ diagram of the pulsars discovered (black dots) and redetected (red circles) by the LOTAAS survey and the known pulsar population (grey dots). The grey dotted and dashed diagonals represent the characteristic ages and the spin-down luminosities, respectively. The solid lines are deathlines from  \citet{Chen93} and \citet{Zhang00}.}
    \label{fig:ppdot}
\end{figure}

When plotted in a $P-\dot{P}$-diagram, Fig.\,\ref{fig:ppdot}, the LOTAAS discoveries have higher characteristic ages $\tau_\mathrm{c}$, lower spin-down luminosities $\dot{E}$ and are generally located closer to the pulsar deathlines where pulsar emission ceases \citep[e.g.][]{Chen93,Zhang00}. The cause of these offsets in spin period and spin period derivative is not immediately obvious, though \citet{Sanidas19} remarked that the 3600\,s integration times of LOTAAS, significantly longer than those of many other pulsar surveys, will increase its sensitivity towards longer period pulsars. 

The offsets may also be related to the low observing frequency of the LOTAAS survey compared to previous surveys, making LOTAAS more sensitive to pulsars with steeper radio spectra. \citet{Jankowski18} found strong correlations of the spectral index $\alpha$ with the spin-down luminosity $\dot{E}$, magnetic field at the light-cylinder $B_\mathrm{LC}$, the spin frequency derivative $\dot{\nu}$, and the characteristic age $\tau_\mathrm{c}$. They found the maximum Spearman correlation coefficient between $\alpha$ values of 276 pulsars and a parameterisation of their spin frequency $\nu$ and $\dot{\nu}$ in the form of $\log_{10} \left(\nu^a|\dot{\nu}|^b\right)$ for values which have $b/a=0.55$. Converting this parameterisation from $\nu$ and $\dot{\nu}$ to $P$ and $\dot{P}$ using $\log_{10} \left(\nu^a|\dot{\nu}|^b\right) \propto \log_{10} \left(P^r\dot{P}^q\right)$, we found that the $b/a=0.55$ correlation corresponds to $r/q=-0.26$ (as $r/q=(-a/b-2)^{-1}$). This indicates that for 1\,dex in $P$, $\dot{P}$ decreases by 0.26\,dex in the $P-\dot{P}$ diagram of Fig.\,\ref{fig:ppdot}, and that pulsars with steeper radio spectra generally have longer spin periods and lower spin period derivatives. 

Longer spin periods and lower spin period derivatives are also seen for pulsars recently discovered with the GBNCC survey \citep{Lynch18} as well as those from the SUPERB survey \citep{Spiewak20} with Parkes. While the GBNCC survey observing at 350\,MHz will also be biased towards pulsars with steeper radio spectra, the SUPERB survey, observing at 1382\,MHz \citep{Keane18} will not, as the majority of pulsars making up the known pulsar population were discovered at those observing frequencies \citep{Edwards01,Manchester01,Cordes06,Keith10}. For the GBNCC survey, population synthesis modelling by \citet{McEwen20} revealed that this bias is introduced through the GBNCC's reduced sensitivity towards pulsars at low Galactic latitudes, due to increased sky temperatures and scattering, while young radio pulsars have not had enough time to move away from their birthplaces in the Galactic plane \citep{FaucherGiguere06}. Hence, the GBNCC survey is biased towards finding pulsars at higher Galactic latitudes, which will predominantly be older pulsars. This is also the case for the SUPERB discoveries as this survey specifically targets higher Galactic latitudes ($|b|>15\degr$; \citealt{Keane18}), as well as LOTAAS, which has worse sensitivity in the Galactic plane compared to the GBNCC survey due to increased sky temperatures and the effects of scattering.

We consider it likely that all three effects combined, sensitivity to longer spin periods, steeper radio spectra and the reduced sensitivity towards young pulsars in the Galactic plane, play a role in discovering pulsars with longer spin periods and lower spin down rates. However, we can not rule out that older pulsars have, on average, steeper radio spectra. Pulsar population synthesis modelling would be required to estimate the impact of each of these effects separately and separate those from any intrinsic pulsar properties, such as a correlation between age and the steepness of the pulsars radio spectra. This task goes beyond the scope of this paper, as these simulations would benefit from using new and updated models of the Galactic electron distribution, the global sky temperature, and scattering \citep[e.g.][]{Yao17,Price21,Geyer17} to properly model the sensitivity for the low observing frequencies of the LOTAAS survey.

\begin{acknowledgements}
    The paper is based (in part) on data obtained with the International LOFAR Telescope (ILT) under project codes LC1\_003, LT3\_001, LC4\_004, LT5\_003, LT5\_004, LC9\_021, LC9\_041, LT10\_015 and LT14\_005. LOFAR \citep{vanHaarlem13} is the Low Frequency Array designed and constructed by ASTRON. It has observing, data processing, and data storage facilities in several countries, that are owned by various parties (each with their own funding sources), and that are collectively operated by the ILT foundation under a joint scientific policy. The ILT resources have benefitted from the following recent major funding sources: CNRS-INSU, Observatoire de Paris and Universit\'e d'Orl\'eans, France; BMBF, MIWF-NRW, MPG, Germany; Science Foundation Ireland (SFI), Department of Business, Enterprise and Innovation (DBEI), Ireland; NWO, The Netherlands; The Science and Technology Facilities Council, UK. We acknowledge the use of the Nan\c cay Data Center computing facility (CDN – Centre de Données de Nançay). The CDN is hosted by the Nançay Radioastronomy Observatory (ORN) in partnership with Observatoire de Paris, Universit\'{e} d’Orl\'{e}ans, OSUC and the CNRS. The CDN is supported by the R\'{e}gion Centre-Val de Loire, d\'{e}partement du Cher. The Nan\c cay Radio Observatory is operated by the Paris Observatory, associated with the French Centre National de la Recherche Scientifique (CNRS). This research was made possible by support from the Dutch National Science Agenda, NWA Startimpuls – 400.17.608. CGB, JWTH, VIK, and DM acknowledge support for this work from the European Research Council under the European Union's Seventh Framework Programme (FP/2007-2013) / ERC Grant Agreement nr. 337062. 
\end{acknowledgements}

\bibliographystyle{aa}
\bibliography{LOTAAS}

\end{document}